\documentclass[11pt]{article}
\usepackage{graphicx}
\usepackage{epsfig}
\usepackage{bm} 

\usepackage{amsfonts,amssymb,amsmath}
\usepackage{latexsym}
\usepackage[english]{babel}

\newcommand{\dd}{{\rm d}}

\newcommand{\eq}{\begin{equation}}
\newcommand{\eeq}{\end{equation}}
\newcommand{\eqn}{\begin{eqnarray}}
\newcommand{\eeqn}{\end{eqnarray}}
\newcommand{\arr}{\begin{eqnarray*}}
\newcommand{\earr}{\end{eqnarray*}}

\newcommand{\eg}{{\it e.g.,}\ }
\newcommand{\ie}{{i.e.,}\ } 

\newcommand{\lp}{\left(}
\newcommand{\rp}{\right)}

\begin{document}

\setlength{\unitlength}{1mm}

\thispagestyle{empty}
 \vspace*{2cm}

\begin{center}
{\bf \LARGE  Kerr-CFT and gravitational perturbations}\\

\vspace*{2.5cm}

{\bf \'Oscar J.C.~Dias,} {\bf Harvey S. Reall,} {\bf Jorge E.~Santos}

\vspace*{1cm}

{\it DAMTP, Centre for Mathematical Sciences, University of Cambridge,\\
 Wilberforce Road, Cambridge CB3 0WA, United Kingdom}\\[.3em]

\vspace*{0.5cm} {\tt O.Dias@damtp.cam.ac.uk, hsr1000@cam.ac.uk,
jss55@cam.ac.uk}

 \vspace{0.5cm} {June 12, 2009}
\end{center}

\begin{abstract}
Motivated by the Kerr-CFT conjecture, we investigate perturbations
of the  near-horizon extreme Kerr spacetime. The Teukolsky equation
for a massless field of arbitrary spin is solved. Solutions fall
into two classes: normal modes and traveling waves. Imposing
suitable (outgoing) boundary conditions, we find that there are no
unstable modes. The explicit form of metric perturbations is
obtained using the Hertz potential formalism, and compared with the
Kerr-CFT boundary conditions. The energy and angular momentum
associated with scalar field and gravitational normal modes are
calculated. The energy is positive in all cases. The behaviour of
second order perturbations is discussed.
\end{abstract}

\noindent


 \setcounter{page}{0} \setcounter{footnote}{0}
\newpage

\tableofcontents


\setcounter{equation}{0}\section{Introduction}

Some time ago, Bardeen and Horowitz (BH) showed that one can take a
near-horizon  limit of the extreme Kerr geometry to obtain a
spacetime similar to $AdS_2 \times S^2$ \cite{Bardeen:1999px}. This
near-horizon extreme Kerr (NHEK) geometry has an $SL(2,R) \times
U(1)$ isometry group, where the $U(1)$ is inherited from the
axisymmetry of the Kerr solution and the $SL(2,R)$ extends the Kerr
time-translation symmetry. Recently, Guica, Hartman, Song and
Strominger (GHSS) have conjectured that quantum gravity in the NHEK
geometry with certain boundary conditions is equivalent to a chiral
conformal field theory (CFT) in 1+1 dimensions \cite{Guica:2008mu}.
Using this, they gave a statistical calculation of the entropy of an
extreme Kerr black hole.

More precisely, GHSS showed that there exist boundary conditions on
the  asymptotic behaviour of the metric such that the asymptotic
symmetry group is generated by time translations plus a single copy
of the Virasoro algebra, the latter extending the $U(1)$ symmetry of
the background. Hence, if a consistent theory of quantum gravity can
be defined in NHEK with these boundary conditions then it must be a
chiral CFT. There has been considerable interest in extending the
Kerr-CFT conjecture,  and entropy calculation, to other extremal
black holes \cite{Lu:2008jk}.

The GHSS boundary conditions are unusual in two respects. First,
they  specify the rate at which components of $h_{\mu\nu}$ (the
deviation of the metric from the NHEK geometry) should behave
asymptotically. We shall refer to these as the ``fall-off"
conditions. Most components decay relative to the background but
some are allowed to be ${\cal O}(1)$ relative to the background.
Secondly, GHSS impose a supplementary boundary condition, namely
that the energy (the conserved charge associated with the generator
$L_0$ of $SL(2,R)$) should vanish.

One motivation for this paper is that the GHSS fall-off conditions
are motivated entirely by considerations of the asymptotic symmetry
group. However, boundary conditions are also required for classical
physics to be predictable from initial data in a non-globally
hyperbolic spacetime such as NHEK (or anti-de Sitter). It is not
clear whether these boundary conditions will be compatible with the
unusual GHSS boundary conditions. Indeed, it is not even clear
whether  the GHSS boundary conditions allow propagating
gravitational degrees of freedom, or whether they lead to physics
similar to Einstein gravity in $AdS_3$, where non-trivial physics is
associated with large gauge transformations (i.e., non-trivial
elements of the asymptotic symmetry group) and black holes that are
locally, but not globally, gauge \cite{btz}. We shall investigate
these issues by studying linearized gravitational perturbations of
NHEK.

Another motivation for studying perturbations of NHEK is associated
with positivity of the energy. The GHSS ``zero energy" condition
arises from the desire to consider only the ground states
corresponding to an extreme Kerr black hole, rather non-extremal
excitations. However, this presupposes that the energy must be
non-negative. The NHEK geometry possesses an ergoregion, inherited
from the ergoregion of the Kerr black hole. It is well-known that,
in the presence of an ergoregion, one can construct initial data for
test matter fields for which the energy of these fields is negative
\cite{friedman}. For a Kerr black hole, this is not a problem
because the positive energy theorem ~\cite{positiveenergy} ensures
that the total energy of the spacetime (black hole plus matter) is
non-negative. This is a non-trivial result, which may not extend to
NHEK.\footnote{ If one wanted to prove such a theorem using
spinorial methods then NHEK would have to admit a spinor field
covariantly constant with respect to some connection. As far as we
know, no such spinor field has been constructed.} Furthermore, in a
spacetime with an ergoregion but no event horizon, e.g. NHEK
(adopting the global perspective), if one imposes boundary
conditions such that there is no energy in the matter fields
entering from infinity, then the total energy of these fields can
only decrease. If it is initially negative then it will become more
negative, suggesting an instability \cite{friedman}.

It should be noted that the issue of NHEK stability is subtle:  BH
pointed out that the singularity theorems imply that there exist
small perturbations of NHEK that will lead to the formation of a
singularity. In this sense, NHEK is unstable. However, as BH also
observed, such a singularity might be hidden inside a tiny black
hole.\footnote{The same might be true in $AdS_d$ for $d \ge 4$:
$AdS_d$ is like a confining box, and a small gravitational
perturbation in a box might be expected to evolve ergodically. If
so, eventually sufficient energy will be concentrated into a small
enough region to produce a tiny black hole. We thank G. Horowitz for
discussion of this point.} If this has positive mass then there
would not be a problem. However, if the energy is negative, or the
singularity is naked, then it would be difficult to make sense of
NHEK.

The NHEK geometry shares many similarities with $AdS_3$: indeed, it
is foliated by warped $AdS_3$ submanifolds, which have been
discussed extensively in recent work on topologically massive
gravity (TMG) \cite{Deser:1982vy}. In TMG, there are propagating
gravitational degrees of freedom but some of these turn out to have
negative energy, signaling a potential instability of $AdS_3$
\cite{Li:2008dq}. In the chiral limit, the  propagating modes are
eliminated by boundary conditions at infinity
\cite{Li:2008dq,Strominger:2008dp}, leaving only pure gauge modes
and BTZ black holes, just as in Einstein gravity . Away from the
chiral limit, $AdS_3$ is unstable but there exist warped $AdS_3$
solutions that might provide an alternative ground state \cite{warped}.
The stability of some of these has been investigated recently
\cite{Anninos:2009zi}. Again, there are propagating modes with
negative energy but these are excluded by boundary conditions.

We now describe the approach we shall take. NHEK is a type D vacuum
spacetime so one can obtain decoupled equations describing
gravitational perturbations using Teukolsky's method
\cite{Teukolsky:1972my,Teukolsky:1973ha}. The Teukolsky equation
turns out to be very similar to the equation governing a massless
scalar field in NHEK, which was discussed by BH, and the qualitative
features of our solutions closely resemble theirs.

By expanding in (spin-weighted, spheroidal) harmonics on the $S^2$
of the NHEK geometry, we reduce the Teukolsky equation to the
equation of a charged massive scalar in $AdS_2$ with a homogeneous
electric field. This equation can be solved in terms of
hypergeometric functions. Depending on the labels $(l,m)$ of the
spheroidal harmonics, the solutions either grow or decay as powers
of the $AdS_2$ radial coordinate, or they are oscillating at
infinity. In the former case, the natural ``normalizable" boundary
conditions lead to quantized frequencies: we shall refer to these as
normal modes. These modes fill out highest-weight representations of
a Virasoro algebra which extends the $SL(2,R)$ isometry group of
$AdS_2$, indeed such modes have been obtained previously in the
context of a charged scalar in $AdS_2$ with electric field
\cite{Strominger:1998yg}. A particularly important set of
normalizable modes are those arising from axisymmetric ($m=0$)
perturbations of NHEK.

The other set of modes are those that oscillate at infinity.
Following BH, we refer to these as traveling waves. These modes
typically have large $m$ for given $l$: $|m| \approx l$. From the
$AdS_2$ perspective, these correspond to modes that have complex
weight with respect to the generator $L_0$ of $SL(2,R)$ and so would
not normally be considered. However, in NHEK it would be very
restrictive to discard these modes since that would correspond to a
restriction on the allowed values of $(l,m)$. Even if such a
restriction were imposed at the linearized level, it would be
violated at the nonlinear level through interactions between modes.

The traveling waves carry energy and angular momentum to infinity.
BH showed that such modes are  associated with superradiant
scattering in the NHEK geometry. However, rather than considering
scattering, we are interested in the question of what happens to
localized initial data. We therefore impose purely outgoing boundary
conditions at infinity. We find that the modes corresponding to
traveling waves become exponentially damped, i.e., they are
quasinormal modes of NHEK, describing the decay of a small
perturbation via radiation to infinity. Therefore NHEK is stable
against linearized gravitational perturbations. The reason that the
above argument for instability based on the energy in matter (or
linearized gravitational) fields fails is that some outgoing waves
carry negative energy to infinity. Hence the energy flux through
infinity need not be positive and so the energy need not decrease
with time.

So far, our discussion of gravitational perturbations has been based
entirely on the Teukolsky equation. However, in order to calculate
the energy, or discuss fall-off conditions on the metric, we need to
know the perturbed metric tensor rather than just the Teukolsky
scalars. Fortunately, there exists a method for determining the
metric perturbation in terms of a scalar potential, called the Hertz
potential \cite{Cohen:1975}-\cite{Stewart1979}. This satisfies an
equation closely related to the Teukolsky equation. Using this, we
obtain explicit results for the form of the metric perturbation.

We find that most (but not quite all) normal modes satisfy the
GHSS fall-off conditions but  traveling waves violate these
conditions. Although one can construct localized wavepackets
involving the latter, they will eventually propagate to infinity and
violate the fall-off conditions. Therefore, at the linearized level,
they should be excluded, leaving just the normal modes.

Next, we consider the energy of the normal modes. To warm-up, we
start  by considering a massless scalar field. We are able to show
that an arbitrary superposition of normal modes has positive energy.
Then we turn to gravitational perturbations. We define the energy of
the latter in the usual way using the Landau-Lifshitz
``pseudotensor". Since the metric perturbation involves second
derivatives of the Hertz potential, the energy involves an integral
of a complicated quantity sixth order in derivatives. Nevertheless,
using a combination of analytical and numerical methods, we find
that the energy of gravitational normal modes is positive, thus
supporting the validity of the GHSS zero-energy condition.

This positive energy result is satisfying but the exclusion of  the
traveling waves is worrying. First, it is worrying that we can
construct initial data that satisfy the fall-off conditions, but
violate these conditions when evolved. It suggests that the initial
value problem, at least for linearized fields, may not be
well-posed. Furthermore, if one goes beyond linearized theory then
interactions between modes will excite traveling waves even if they
are not present initially.\footnote{ The only way to escape this
conclusion is to consider only axisymmetric ($m=0$) modes, which
form a consistent truncation of the full set of modes.} So one might
worry about well-posedness of the nonlinear theory too. It is
possible that these problems are cured by backreaction, i.e,  going
beyond the linearized approximation. We shall discuss this further
at the end of the paper.

This paper is organized as follows. In section \ref{sec:teuk}, we
derive  and solve the Teukolsky equation in the NHEK background,
obtaining the spectrum of normal, and quasinormal modes. In section
\ref{sec:metric} we introduce the Hertz potential and use it to
obtain the explicit form of linearized perturbations. We compare the
asymptotic behaviour of these with the GHSS boundary conditions. We
then calculate the energy of scalar field and gravitational normal
modes. Finally, section \ref{sec:discussion} discusses how going
beyond the linearized approximation may solve some of the problems
just discussed.

 {\bf Note added.} As this work was nearing completion, we learned
that another group is exploring similar issues \cite{santabarbara}.

\setcounter{equation}{0}
\section{Massless fields of arbitrary spin in NHEK
\label{sec:teuk}}

\subsection{NHEK and its
Newman-Penrose tetrad \label{sec:NHkerr}}

In global coordinates the NHEK metric is \cite{Bardeen:1999px} (we
use  the notation of Ref. \cite{Guica:2008mu} and, because we shall
employ the Newman-Penrose formalism, a negative signature metric)
\eq ds^2= -2G J\Omega^2(\theta)\lp -(1+r^2)dt^2 +\frac{dr^2}{1+r^2}
+d\theta^2+\Lambda^2(\theta)(d\phi +r dt)^2\rp , \label{NHkerr} \eeq
with
\eq
 \Omega^2(\theta)\equiv \frac{1}{2}(1+\cos^2\theta), \qquad
 \Lambda(\theta)=\frac{2\sin\theta}{1+\cos^2\theta},\qquad GJ=G^2M^2_{\rm
ADM}\equiv M^2.
  \label{NHkerrAux} \eeq
Surfaces of constant $\theta$ are warped $AdS_3$ geometries, \ie a
circle fibred over $AdS_2$ with warping parameter
$\Lambda^2(\theta)$. The isometry group is $SL(2,R) \times U(1)$. BH
showed that the solution is geodesically complete, with timelike
infinities at $r =\pm \infty$. There is an ergoregion (where
$\partial/\partial t$ is spacelike) which extends to $r=\pm
\infty$.

In the next subsection we study perturbations in the NHEK using the
Teukolsky formulation. For that we need the Newman-Penrose (NP)
tetrad, spin coefficients and directional derivatives. In Appendix
\ref{sec:geodesic} we obtain the shear-free null geodesics of this
background and use them to construct the associated NP null tetrad
\cite{ChandrasekharBook}, ${\bf e}_{(1)}= \bm{\ell}$, ${\bf
e}_{(2)}={\bf n}$, ${\bf e}_{(3)}={\bf m}$, ${\bf e}_{(4)}={\bf
m^*}$, where (coordinates are listed in the order $\{t,r,\theta,\phi \}$)
\eqn
 && \ell^\mu=\frac{1}{1+r^2}\lp 1, 1+r^2,0,-r\rp ,\nonumber\\
 && n^\mu=\frac{1}{4M^2\Omega^2(\theta)}\lp 1, -(1+r^2),0,-r\rp ,\nonumber\\
 && m^\mu=\frac{1}{\sqrt{2}\,M(1+ i \cos\theta)}\lp 0,0,1,i
\Lambda^{-1}(\theta) \rp ,
 \label{NPtetrad}
\eeqn
and ${\bf e}^{(a)}=\eta^{(a)(b)}{\bf e}_{(b)}$ with non-vanishing
symmetric $\eta^{(a)(b)}=\eta_{(a)(b)}$ given by
$\eta^{(1)(2)}=-\eta^{(3)(4)}=1$. This NP tetrad satisfies the
normalization and orthogonality conditions \eqref{NPnormOrth}, and
the null vector $\bm{\ell}$ is tangent to affinely parametrized geodesics:
$\ell^\mu \nabla_\mu \ell_\nu =0$.

The unperturbed Weyl scalars in the NHEK geometry are computed using
\eqref{WeylScalars}, yielding
\begin{eqnarray}
&& \Psi_0 = \Psi_1 = \Psi_3 = \Psi_4 \equiv 0\,, \nonumber \\
&&\Psi_2 = -\left[ M^2 (1- i \cos\theta)^3\right]^{-1}.
\label{WeylScalarsNH}
\end{eqnarray}
The first line confirms that this solution is indeed Petrov type D.

\subsection{Teukolsky master equation  \label{sec:TeukolskyHertz}}

Teukolsky has shown how, for type D spacetimes, one can use the NP
formalism to derive a system of decoupled equations, that
furthermore separate into an angular and radial part, for the
perturbations of several NP scalars
\cite{Teukolsky:1972my,Teukolsky:1973ha}. For gravitational
perturbations,  the relevant quantities are the perturbed Weyl
scalars $\Psi_0^{(1)}$ (spin $s=+2$) and $\Psi_4^{(1)}$ ($s=-2$);
the complex NP scalars $\phi_{0,1}$ for spin $s=\pm 1$ Maxwell
perturbations; the Weyl fermionic scalars $\chi_{0,1}$ for massless
spin $s=\pm \frac{1}{2}$ perturbations; and the scalar field $\Phi$
for massless spin $s=0$ perturbations. Teukolsky's master equation
encompasses all of these cases \cite{Teukolsky:1973ha}.


Using the NP quantities listed in appendix \ref{sec:geodesic}, we
find that  the Teukolsky master equation for spin $s$ field
perturbations $\Psi^{(s)}$ in the NHEK geometry is
\begin{eqnarray}
&& \hspace{-1cm} \frac{1}{\left(1+r^2\right)}\partial_t^2\Psi^{(s)}
-\frac{2r}{\left(1+r^2\right)} \partial_t\partial_\phi \Psi^{(s)} +
\left(\frac{r^2}{1+r^2}-\frac{\left(
1+\cos^2\theta\right)^2}{4\sin^2\theta}
\right)\partial_\phi^2\Psi^{(s)}
 \nonumber \\
&& \hspace{0cm}  -\left(1+r^2\right)^{-s}\partial_r
\left(\left(1+r^2\right)^{s+1}\partial_r  \Psi^{(s)} \right)
-\frac{1}{\sin\theta}\,\partial_\theta \left(\sin\theta \,
\partial_\theta \Psi^{(s)} \right)-2s\frac{r}{\left(1+r^2\right)} \partial_t\Psi^{(s)} \nonumber \\
&& \hspace{1cm}
 -2s\left(\frac{1 } {\left(1+r^2\right) }+ i
\frac{\cos\theta}{\sin^2\theta}+i\frac{1}{2} \cos\theta \right)
\partial_\phi  \Psi^{(s)} +\left(s^2
\cot^2\theta-s\right)  \Psi^{(s)} = T_{(s)} \,.
  \label{MasterEq}
\end{eqnarray}
We have allowed for the possibility of a source term on the RHS (see
Appendix \ref{sec:Wald}). The relation between the nomenclature used
here and the original notation of Teukolsky \cite{Teukolsky:1973ha}
is $\{ \Psi^{(2)},\Psi^{(1)},\Psi^{(1/2)} \}=\{ \Psi_0,\phi_0,\chi_0
\}$ and $\{ T_{(2)},T_{(1)},T_{(1/2)}\}=\{ T_0, J_0,T_{\chi_0}\}$
for positive spin. For negative spin the map is
 $\{ \Psi^{(-2)},\Psi^{(-1)},\Psi^{(-1/2)}
\} =\{ \lp -\Psi_{2}\rp^{\frac{4}{3}}\Psi_4,\lp
-\Psi_{2}\rp^{\frac{2}{3}}\phi_2,\lp
-\Psi_{2}\rp^{\frac{1}{3}}\chi_1 \}$. Here, the powers of the
unperturbed Weyl scalar $\Psi_{2}$ are those that allow for the
separation of the master equation, when we further assume an ansatz
for the perturbation that is a radial function times the
spin-weighted spheroidal harmonic; see \eqref{TeukAnsatz}. For the
source term one has the map $\{ T_{(-2)},T_{(-1)},T_{(-1/2)} \}=\{
T_4, J_2,T_{\chi_1}\}$. These relations are summarized in
Table \ref{Table:Teuk}.

\begin{table}[ht]
\begin{eqnarray}
\nonumber
\begin{array}{||c||||c|c||c|c||c|c||c||}\hline\hline\hline\hline
 \Psi^{(s)}  & \lp -\Psi_{2}\rp^{\frac{4}{3}} \Psi_4 & \Psi_0 &   \lp -\Psi_{2}\rp^{\frac{2}{3}} \phi_2 & \phi_0
 & \lp -\Psi_{2}\rp^{\frac{1}{3}} \chi_1 & \chi_0 &  \Phi \\
\hline \hline
 s & -2 & 2 &  -1 & 1 & -\frac{1}{2} & \frac{1}{2} & 0 \\
 \hline
 T_{(s)}  & 2 \lp-\Psi_{2}\rp^{\frac{4}{3}}T_4 & 2 T_0   & \lp -\Psi_{2}\rp^{\frac{2}{3}}J_2 & J_0
 & \lp -\Psi_{2}\rp^{\frac{1}{3}}T_{\chi_1}& T_{\chi_0} & T_{\Phi} \\ 
\hline\hline\hline\hline
\end{array}
\end{eqnarray}
\caption{Teukolsky fields $\Psi^{(s)}$, spin $s$ and source terms
for the master equation \eqref{MasterEq}.} \label{Table:Teuk}
\end{table}

\subsection{Separation of variables}

We shall solve the Teukolsky equation in the NHEK geometry by separation of variables. Assuming
\begin{equation}
\Psi^{(s)}= \left\{
\begin{array}{ll}
e^{-i\omega t}e^{i m\phi}R^{(s)}_{lm\omega}(r)S^{(s)}_{lm}(\theta)\lp
-\Psi_{2}\rp^{-\frac{2s}{3}} \,, & \qquad s\leq 0 \,,\label{TeukAnsatz} \\
e^{-i\omega t}e^{i m\phi}R^{(s)}_{lm\omega}(r)S^{(s)}_{lm}(\theta) \,, & \qquad
s\geq 0 \,,
\end{array}
\right.
\end{equation}
equation \eqref{MasterEq} separates into an angular and radial
equations. The angular equation is
\eq
 \frac{1}{\sin\theta}\,\frac{d}{d\theta}\lp \sin\theta \frac{d}{d\theta}
S_{lm}^{(s)}(\theta)\rp
 +\left[ (C\,\cos\theta)^2-2 s C \,\cos\theta +s +\Lambda_{lm}^{(s)} -\frac{\lp
m+s\,\cos\theta \rp^2}{\sin^2\theta}\right]S_{lm}^{(s)}(\theta)=0\,,
 \label{AngEq}
\eeq
for $C=m/2$ and where $\Lambda_{lm}^{(s)}$ is the separation
constant. Its eigenfunctions are the spin-weighted spheroidal
harmonics $e^{i m\phi}S_{lm}^{(s)}(\theta)$ (the nomenclature
usually includes an appropriate normalization factor; see \eg
\cite{Press:1973zz}), with positive integer $l$ specifying the
number of zeros, $\ell-{\rm max}\{|m|,|s|\}$, of the eigenfunction.
The associated eigenvalues $\Lambda_{lm}^{(s)}$ can be computed
numerically with very good accuracy and are specified by $s,l,m$
subject to the regularity constraints that $-l\leq m\leq l$ must be
an integer and $l\geq |s|$. The transformation $\theta \rightarrow
\pi-\theta$ can be used to show that
 \eq
 \Lambda_{lm}^{(s)}=\Lambda_{l(-m)}^{(s)}, \qquad \Lambda_{lm}^{(-s)}=\Lambda_{lm}^{(s)}+2s\,. \label{spinLambda}
  \eeq
We also note that, to leading order in $C$,
$\Lambda_{lm}^{(s)}=(l-s)(l+s+1)+\mathcal{O}(C)$. This is useful
when $|m| \ll l$.

Equation \eqref{AngEq} represents the most standard
way to write the spin-weighted spheroidal harmonic equation. However, it will be convenient here to work with shifted eigenvalues
$\widetilde{\Lambda}^{(s)}_{lm}$ defined by
\begin{equation}
\widetilde{\Lambda}^{(s)}_{lm}\equiv
\Lambda_{lm}^{(s)}+s^2+s-7C^2.
\end{equation}
The advantage of using these quantities is that they have the symmetry
\begin{equation}
\widetilde{\Lambda}^{(-s)}_{lm}=\widetilde{\Lambda}^{(s)}_{lm}.\label{NewLambdaSymm}
\end{equation}

Notice that in the Kerr background with mass $M$ and angular
velocity $\Omega_H$ the angular equation for spin $s$ perturbations
is also \eqref{AngEq} but with $C_{\rm Kerr}=a\widetilde{\omega}$,
where $a=2Mr_+ \Omega_H$ is Kerr's rotation parameter and
$\widetilde{\omega}$ the wave's frequency in this geometry. As
observed in \cite{Bardeen:1999px}, in the near-horizon limit of extreme
Kerr, all finite frequencies $\omega$ in the NHEK throat correspond
to the single frequency $\widetilde{\omega}=m\Omega_H^{\rm
ext}=\frac{m}{2M}$ in the extreme Kerr geometry. This
$\widetilde{\omega}$ corresponds precisely to the marginally
unstable superradiant frequency, and in the NH limit one finds
$C_{\rm Kerr}=M \widetilde{\omega} \rightarrow C=m/2$.

Writing for any spin,
 \eq
 R^{(s)}_{lm\omega}(r)=(1+r^2)^{-s/2}\,\Phi^{(s)}_{lm\omega}(r)\,,
 \label{RadialAns}
 \eeq
we find that the radial equation associated with \eqref{MasterEq}
can be written also in a unified way as
 \eq
 \frac{d}{dr}\left[ (1+r^2) \frac{d}{dr} \Phi^{(s)}_{lm\omega}(r)
\right] - \left[ \mu^2 - \frac{(\omega + q r)^2}{1+r^2} \right]
\Phi^{(s)}_{lm\omega}(r) = 0 \,, \label{RadialEq}
 \eeq
with
\begin{eqnarray}
&& q = m-is \,,
\nonumber \\
&& \mu^2 = q^2 + \widetilde{\Lambda}^{(s)}_{lm}
 =
\Lambda_{lm}^{(s)}+s-2 i s m-\frac{3m^2}{4} \,.
\label{ChargeMassAdS}
\end{eqnarray}
\begin{figure}[t]
\centerline{\includegraphics[width=.30\textwidth]{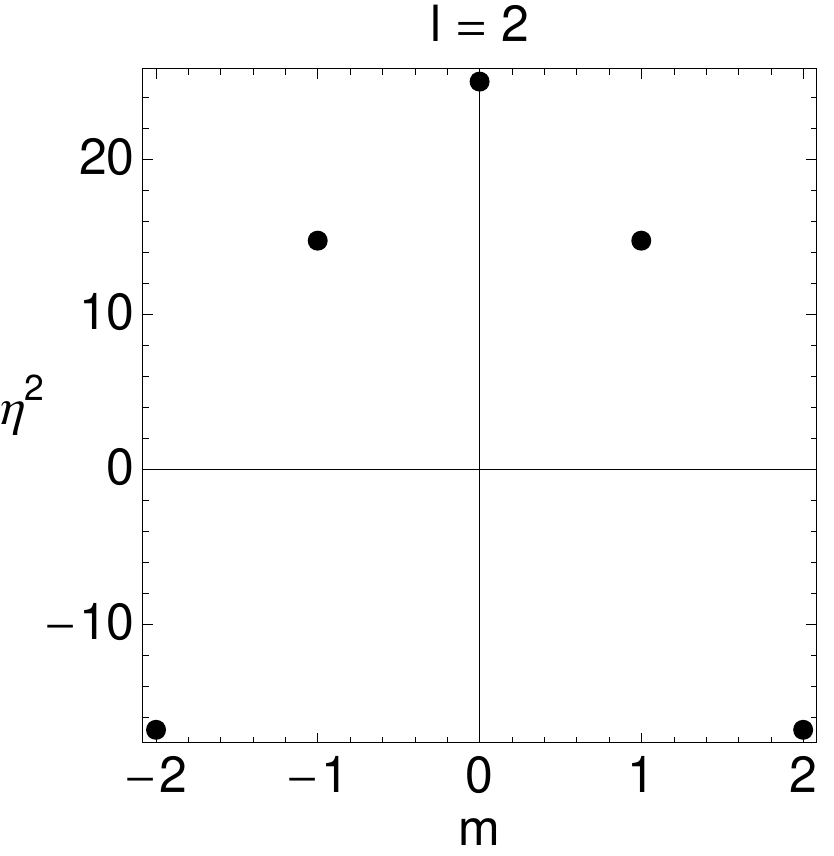}
\hspace{2cm}\includegraphics[width=.30\textwidth]{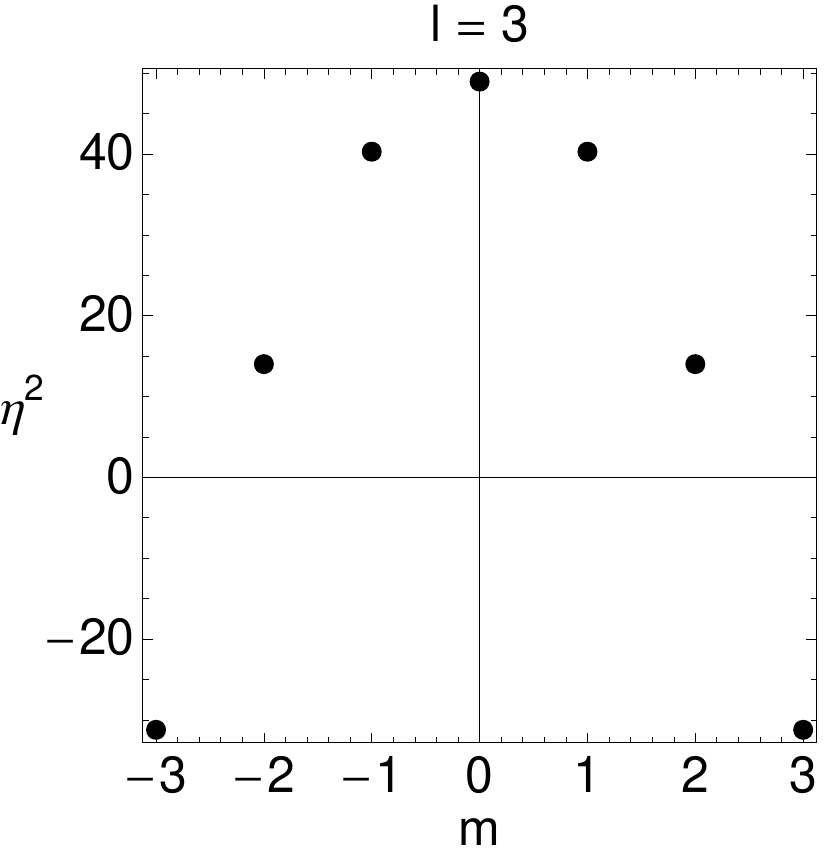}}
\caption{Values of $\eta^2$, defined in \eqref{eqn:etadef}, for
$|s|=2$, and a) $l=2$ and b) $l=3$. } \label{fig:Eigenvalues}
\end{figure}
\begin{figure}[t]
\centerline{\includegraphics[width=.30\textwidth]{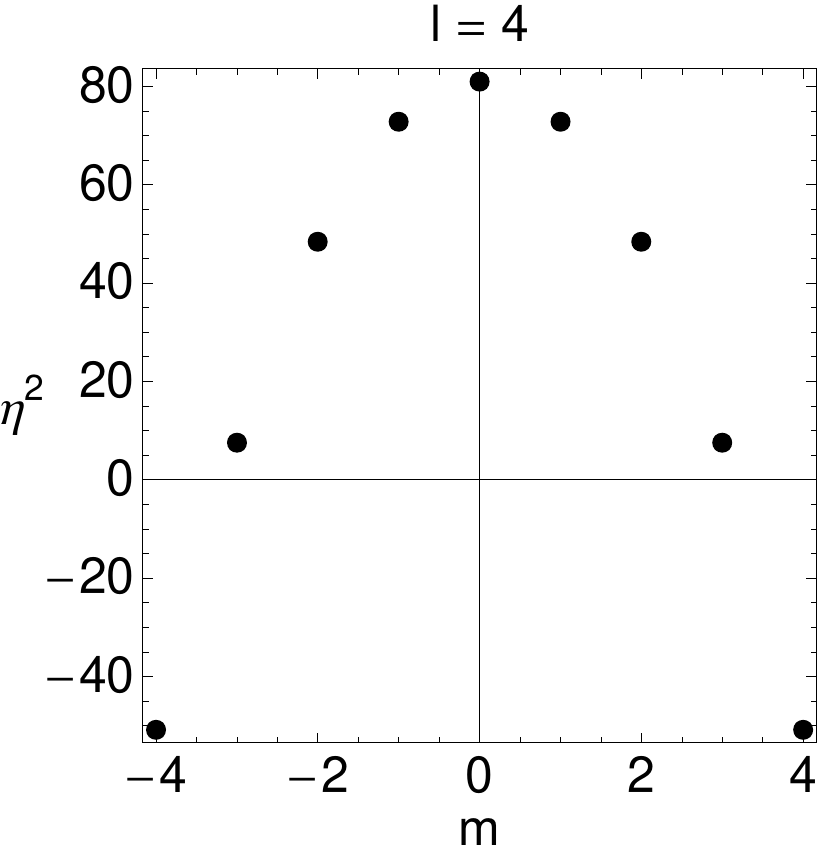}
\hspace{2cm}\includegraphics[width=.30\textwidth]{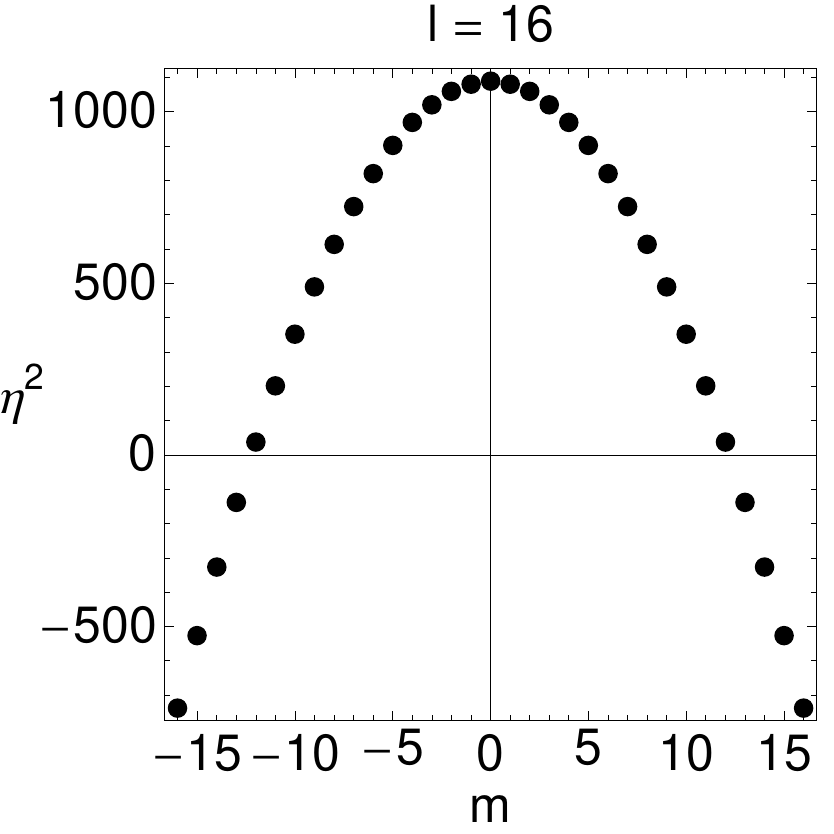}}
\caption{Values of $\eta^2$, defined in \eqref{eqn:etadef}, for
$|s|=2$, and a) $l=4$ and b) $l=16$. } \label{fig:Eigenvalues2}
\end{figure}
This is exactly the equation for a charged massive scalar field in
$AdS_2$ with a homogeneous electric field: take the $AdS_2$ metric
in global coordinates,
 \eq
 ds_2^2 =(1+r^2) dt^2 - \frac{dr^2}{1+r^2} dr^2\,,
 \eeq
 and the electric field to arise from the potential
 \eq
 A = r dt.
 \eeq
Define the covariant derivative for a field of charge $q$ as
 \eq
 {\cal D} = \nabla - i q A,
 \eeq
where $\nabla$ is the Levi-Civita connection in $AdS_2$. The equation for a charged scalar field $\Phi(t,r)$ with mass $\mu$
 is then
  \eq
 {\cal D}^2 \Phi + \mu^2 \Phi=0\,.
 \eeq
Assuming
 \eq
 \Phi(t,r) = e^{-i\omega t}\,\Phi(r)\,,
 \eeq
the equation of motion reduces to \eqref{RadialEq}. Therefore, a
general spin $s$ perturbation with angular momentum $m$ in NHEK
obeys  the wave equation for a massive charged scalar field in
$AdS_2$ with a homogeneous electric field. However, note that the
charge $q$ is complex, as is the squared mass $\mu^2$, although
$\mu^2-q^2$ is real. The problem of a massive charge scalar field in
$AdS_2$ with homogeneous electric field was studied in Ref.
\cite{Strominger:1998yg}, where solutions corresponding to highest
weight representations of a Virasoro algebra extending $SL(2,R)$
were obtained. We shall recover the same solutions in the next
section.

\subsection{Solving the radial equation}

Asymptotically, the solutions of \eqref{RadialEq} behave as
\begin{equation}
\Phi(r) \sim |r|^{-1/2 \pm \eta/2},
\label{Phifalloff}
\end{equation}
where
\begin{equation}
 \label{eqn:etadef}
 \qquad \eta =  \sqrt{1+4\lp \mu^2-q^2 \rp}=\sqrt{1+4 \widetilde{\Lambda}^{(s)}_{lm}}, \qquad {\rm Im}(\eta) \ge 0.
\end{equation}
Note that $\eta(s,l,m) = \eta(-s,l,m)=\eta(s,l,-m)$. We can now see that the modes can exhibit qualitatively different
behaviour,  depending on the value of $(l,m)$, as first noticed by
BH (for $s=0$). Some modes have real $\eta$ and others have
imaginary $\eta$. For example, axisymmetric modes ($m=0$), have, for
general $s$,
\begin{equation}
 \eta = 2 l + 1 \qquad (m=0).
\end{equation}
\ie such modes exhibit power-law behaviour at infinity. However, for
certain other modes, specifically those with $|m| \approx l$,
$\eta$ is imaginary and hence the solutions
oscillate at infinity. In Figs. \ref{fig:Eigenvalues} and
\ref{fig:Eigenvalues2} we show how $\eta^2$ depends on $m$ for
gravitational perturbations with some different values of $l$.

It is interesting to ask which modes have the smallest real value
for $\eta$ since these will give the normal modes that decay
most slowly at infinity. For gravitational perturbations ($|s|=2$)
we have calculated $\eta$ for all $(l,m)$ with $l \le 30$ and find
that the mode with the smallest real value for $\eta$ occurs for
$l=4$, $|m|=3$, which gives $\eta=2.74$.

Equation \eqref{RadialEq} can be solved exactly. This is not a
surprise since in the Kerr geometry, Teukolsky and Press
\cite{Teukolsky:1974yv} found that the corresponding Teukolsky
radial equation can also be analytically solved in the particular
case where we have extreme Kerr and a wave frequency that saturates
the superradiant bound, $\widetilde{\omega}=m\Omega_H^{\rm ext}$. As
discussed after \eqref{NewLambdaSymm}, all frequencies in NHEK
correspond to the single superradiant threshold frequency in the
extreme Kerr. So we indeed expect this property for the radial
equation in NHEK.

Introducing the new radial coordinate,
 \eq
z=\frac{1}{2}\lp 1-i r\rp \,,
 \label{NewRadCoord}
\eeq
and redefining the radial wavefunction as
\eq
 \Phi^{(s)}_{lm\omega}(r)= z^{\alpha} (1-z)^{\beta}\,F \,, \qquad {\rm
with} \quad \alpha\equiv \frac{1}{2}\lp \omega -iq \rp \,,\qquad
\beta\equiv \frac{1}{2}\lp \omega+iq \rp  \,,
 \label{WaveFunc}
 \eeq
the radial equation \eqref{RadialEq} can be rewritten as
\begin{equation}
 z(1-z)\partial_z^2 F+ \left[ 2\alpha -2\lp \alpha+\beta \rp z
 \right]
\partial_z F - \left[ (\alpha+\beta+1)(\alpha+\beta-2)+\lp q^2-\mu^2+2 \rp\right ]
F=0\,.
 \label{RadialEq2}
\end{equation}
This wave equation is a standard hypergeometric equation
\cite{abramowitz}, $z(1-z)\partial_z^2 F+[c-(a+b+1)z]\partial_z F-ab
F=0$, with
\begin{equation}
 a=\frac{1}{2}\lp 1+\eta+2\omega \rp \,, \qquad
b=\frac{1}{2}\lp 1-\eta+2\omega \rp\,,
 \qquad c= 1+\omega-i q \,,
\end{equation}
and hence the most general solution in the neighborhood of $z=0$ is
\cite{abramowitz}
\begin{equation}
 \Phi^{(s)}_{lm\omega} = A  z^\alpha (1-z)^\beta F(a,b,c,z)+ B z^{\alpha+1-c}(1-z)^\beta
F(a-c+1,b-c+1,2-c,z).
\label{RadialSol}
\end{equation}
We render this function single valued in the complex $z$ plane by
taking  branch cuts to run from $-\infty$ to $0$ and from $1$ to
$+\infty$, corresponding to taking $|\arg(z)|<\pi$, $|\arg(1-z)| <
\pi$.  Note that the branch cuts do not intersect the line ${\rm
Re}(z)=1/2$, which corresponds to real $r$.

\subsection{Boundary conditions}\label{sec:BCs}

The above solution of the radial equation is regular for all finite
$r$. Using standard properties of the hypergeometric function, we
find that  it exhibits the following behaviour as $r \rightarrow \pm
\infty$: \eq
  \Phi^{(s)}_{lm\omega} \approx \Gamma(b-a) C^\pm e^{\pm i \pi(\beta-\alpha -a)/2}
  \left(\frac{|r|}{2}\right)^{-(1+\eta)/2}
  + \Gamma(a-b) D^{\pm} e^{\pm i \pi(\beta-\alpha -b)/2}  \left(\frac{|r|}{2}\right)^{-(1-\eta)/2},\label{RadialSolAsymp}
\eeq
 where
 \begin{eqnarray}
&& C^{\pm} = A\frac{\Gamma(c)}{\Gamma(b) \Gamma(c-a)} - B e^{\pm i
\pi c} \frac{\Gamma(2-c)}{\Gamma(b-c+1)\Gamma(1-a)},
 \nonumber\\
&& D^{\pm} =   A\frac{\Gamma(c)}{\Gamma(a) \Gamma(c-b)} - B e^{\pm i
\pi c} \frac{\Gamma(2-c)}{\Gamma(a-c+1)\Gamma(1-b)}.
\label{amplitudes}\end{eqnarray}
The boundary conditions now depend
on whether $\eta$ is real or imaginary.

\subsubsection{Normal modes}

Assume that $\eta$ is real. In this case, we impose normalizable
boundary conditions, corresponding to demanding that $D^+=D^-=0$, a
pair of simultaneous equations for $A$, $B$. Non-zero solutions
exist only if the determinant of this system vanishes. Using $\Gamma(z)\Gamma(1-z)=\pi/\sin(\pi z)$, this gives
\eq
 \frac{(1-c)\pi}{\Gamma(a) \Gamma(1-b) \Gamma(c-b) \Gamma(a-c+1)} =
 0. \label{normalModeCond}
\eeq This imposes a quantization condition on $\omega$,
corresponding to the  two solutions $a=-n$ and $1-b=-n$ where
$n=0,1,2,\ldots$.\footnote{At first sight, condition
\eqref{normalModeCond} could also be satisfied if we imposed $c=1$,
\ie $\omega=iq$. However, a more careful analysis rules out this
possibility because for $c=1$, \eqref{RadialSol} is not a solution
of the problem: one must allow for a logarithmic dependence in the second part.
Redoing the analysis with the appropriate regular
radial solution for this special case \cite{abramowitz}, we conclude
that nothing physically special occurs for $c=1$.\label{foot}}
 The former solution gives
\eq
 \omega = -(n+1/2+\eta/2), \qquad n=0,1,2,\ldots, \qquad B=0,
\eeq
and the latter gives
\eq
 \omega = n+1/2+\eta/2, \qquad n=0,1,2,\ldots, \qquad A=0.
\eeq
We can summarize the normal mode spectrum as
\eq
 \omega = \pm (n+1/2+\eta/2)\,, \qquad n=0,1,2,\ldots .
\label{FreqSpectrum} \eeq This is precisely the spectrum of normal
modes found for a massive charged  scalar in $AdS_2$ with a
homogeneous electric field in Ref. \cite{Strominger:1998yg}.

Note that we have allowed $\omega$ to be positive or negative. This
is  because the Teukolsky equation for $s \ne 0$ is {\it not}
invariant under complex conjugation, so negative frequency solutions
are not simply related to positive frequency solutions by complex
conjugation, they have to be considered separately. The two possible
signs correspond to the two different helicities of the field. The
radial equation is invariant under $\omega \rightarrow -\omega$, $r
\rightarrow -r$ hence $\Phi^{(s)}_{lm(-\omega)}(r) \propto
\Phi^{(s)}_{lm\omega}(-r)$.

For $n=0$, the positive frequency solution of the radial equation is
\eq
 \Phi^{(s)}_{lm(n=0)} \propto z^{-(1+\eta)/4+iq/2} (1-z)^{-(1+\eta)/4-iq/2},
\eeq
 and the negative frequency solution is obtained by $r \rightarrow -r$, i.e.,
 $z \rightarrow 1-z$. The solutions with positive $n$ are related to
 these $n=0$ solutions by multiplication by a polynomial of degree $n$ in $z$.

\subsubsection{Traveling waves \label{sec:travel}}

Now consider the case of imaginary $\eta$. Define $\tilde{\eta}>0$
by  $\eta = i \tilde{\eta}$. The radial function oscillates at
infinity, corresponding to incoming or outgoing waves (see Appendix
\ref{sec:veloc} for details). Rather than considering scattering in
NHEK, we shall impose boundary conditions corresponding to purely
outgoing waves at infinity, which will discretize the frequency
$\omega$ and render it complex. A solution with positive imaginary
part corresponds to an instability, and a solution with negative
imaginary part is a quasinormal mode.

As discussed by BH, there are two inequivalent notions of
``outgoing" that  one can use in NHEK because the phase velocity and
group velocity of wavepackets need not have the same sign, e.g. for
positive $\omega$ and $m$, the group and phase velocities have the
same sign at $r =+ \infty$ but opposite sign at $r = -\infty$ (see
Table \ref{Table:Veloc}). Physical boundary conditions correspond to
the notion of ``outgoing" defined using the group velocity. However,
it is easier to analyze the case of outgoing phase, so we shall
consider this case first.

Assume that ${\rm Re}(\omega)>0$. Then the solutions with outgoing
phase at $r \rightarrow \pm \infty$ are the solutions  with
$C^\pm=0$. This leads to the quantization condition
 \eq
 \frac{(1-c)\pi}{\Gamma(b) \Gamma(1-a) \Gamma(c-a) \Gamma(b-c+1)} = 0,
\eeq with solution $1-a=-n$, $n=0,1,2,\ldots$ ($b=-n$ is
inconsistent  with ${\rm Re}(\omega)>0$), which gives $\omega =
n+1/2-i\tilde{\eta}/2$. Repeating the exercise for ${\rm
Re}(\omega)<0$ requires $D^\pm=0$, and leads to $\omega =
-(n+1/2)-i\tilde{\eta}/2$. We can summarize the result as \eq
 \omega = n+1/2 - i \tilde{\eta}/2, \qquad n \in \mathbb{Z}
\eeq The imaginary part is negative, hence these are quasinormal
modes.  This is a little surprising. BH pointed out that the energy
flux (for positive frequency modes) has the same sign as the phase
velocity. Hence outgoing phase should correspond to outgoing energy
at infinity. As discussed in the introduction, this is precisely the
situation in which one expects an instability associated with the
negative energy in matter fields within the ergoregion becoming
increasingly negative. We have found that outgoing phase leads to
stable quasinormal modes rather than an instability. However, these
boundary conditions are unphysical: we are arranging that an initial
wavepacket (at finite $r$) composed of modes with positive
$\omega,m$ does not propagate to $r= -\infty$ by sending in an
appropriate (finely tuned) wavepacket from $r= -\infty$ to scatter
with it in such a way as to produce only a wavepacket propagating to
$r = +\infty$. This is analogous to boundary conditions for a Kerr
black hole in which one arranges that initial data leads to no waves
crossing the future horizon by sending in appropriate waves from the
past horizon. Presumably, the fine-tuning is the reason that we do
not see an instability here.

Now consider the physical boundary conditions corresponding to
``outgoing" defined with respect to the group velocity. Assume that
${\rm Re}(\omega)>0$ and $m > 0$. BH showed that, under these
conditions, the phase and group velocities have the same sign for $r
\rightarrow \infty$ but opposite sign for $r \rightarrow -\infty$.
Hence the boundary conditions that we need are $C^+ = D^- = 0$. In fact, the same holds for ${\rm Re}(\omega)<0$ and $m >0$ (see Appendix \ref{sec:veloc}).
Using the identity $\Gamma(z)\Gamma(1-z)=\pi/\sin(\pi z)$, we find
that the quantization condition is \eq
 \sin (\pi b) \sin [\pi (c-a)] e^{-i\pi c} = \sin (\pi a) \sin [\pi (c-b)] e^{i \pi c},
\eeq which gives\footnote{A solution corresponding to $c$ taking integer values is ruled out for the reason discussed in footnote \ref{foot}.}
 \eq
 \omega = n + \frac{1}{2} - \frac{i}{2\pi} \log \left[
 \frac{ \cosh [ \pi (\tilde{\eta}/2+m)]}{\cosh [ \pi (\tilde{\eta}/2-m)]} \right], \qquad \quad n \in \mathbb{Z}, \qquad (m>0)
\eeq
where we have specialized to scalar field ($s=0$) or
gravitational ($\pm 2$) perturbations for simplicity. Repeating the
analysis for  $m <0$ requires $C^-=D^+=0$. The general result is
\eq
 \omega = n + \frac{1}{2} - \frac{i}{2\pi} \log \left[ \frac{ \cosh
 [ \pi (\tilde{\eta}/2+|m|)]}{\cosh [ \pi (\tilde{\eta}/2-|m|)]} \right], \qquad n \in \mathbb{Z}
\eeq
We see that ${\rm Im}(\omega)<0$ hence these are stable quasinormal modes.
So NHEK is stable against linearized gravitational (and scalar field) perturbations.

\section{Metric perturbations\label{sec:metric}}

\subsection{Hertz potentials}

The NP scalar perturbations $\Psi^{(s)}$ are useful because they are
invariant under infinitesimal diffeomorphisms and under rotations of
the NP tetrad. Many physically interesting quantities can be
computed directly from the knowledge of these NP fields
\cite{Teukolsky:1972my,Teukolsky:1973ha,ChandrasekharBook}. However,
in some problems as is our case, we really need to know the
perturbations of the metric itself, $h_{\mu\nu}$, or
the perturbations of the Maxwell or Weyl fermionic vector fields,
respectively $A_\mu$ and $\chi_\mu$. Cohen and Kegeles
\cite{Cohen:1975,Cohen:1979}, and Chrzanowski
\cite{Chrzanowski:1975} have proposed a unique map that provides the
$h_{\mu\nu}$, $A_\mu$ or $\chi_\mu$ perturbations given the
so-called Hertz potential $\Psi_{\rm H}^{(s)}$ (see a good
discussion also in \cite{Stewart1979}).  Wald proved
Cohen-Kegeles$-$Chrzanowski's results \cite{Wald}. See Appendix
\ref{sec:Wald} for a detailed discussion of these works. The main
conclusion is that the Hertz potential also obeys a pair of
decoupled equations, again one for positive and the other for
negative $s$. These are written in equations
\eqref{HertzPotential:NegSpin} and \eqref{HertzPotential:PosSpin}.

For gravitational perturbations, this method yields the metric
perturbation in a particular gauge: the ingoing (outgoing) radiation
gauge IRG (ORG), specified by the conditions
\begin{equation}
\ell^\mu h_{\mu \nu} = g^{\mu\nu}h_{\mu\nu}=0 \: {\rm (IRG)}, \qquad
n^\mu h_{\mu\nu} =  g^{\mu\nu}h_{\mu\nu}=0 \: {\rm (ORG)}.
\label{gauges}
\end{equation}
At first sight, these gauge conditions appear overdetermined but it
has been shown that, for perturbations of a type II vacuum
spacetime, there is a residual gauge freedom that allows one to
impose the IRG provided that $\ell^\mu \ell^\nu T_{\mu\nu}=0$ where
$\ell$ is the repeated principal null direction and $T_{\mu\nu}$ the
stress-tensor of any matter perturbation present
\cite{Price:2006ke}. Similarly, for type D one can impose either the
IRG or the ORG (if $n^\mu n^\nu T_{\mu\nu}=0$). The spin of the
Hertz potential corresponds to these two different gauges: the
metric perturbation in the IRG (ORG) is obtained from the Hertz
potential with $s=-2$ ($s=+2$). The two Hertz potentials contain
exactly the same physical information, so one need only work with
one of them.

For vacuum type D spacetimes, the Hertz potential itself satisfies a
master equation. For the Kerr solution, this master equation turns
out to be exactly the same as for the original NP scalars
$\Psi^{(s)}$, equation \eqref{MasterEq}, with no source term on the
RHS \cite{FrolovNovikov}. We have checked that the same is true for
NHEK. More concretely, $\Psi_{\rm H}^{(s)}=\{ \Psi_{\rm
H}^{(-2)},\Psi_{\rm H}^{(-1)},\Psi_{\rm H}^{(-1/2)} \}$ are the
Hertz potentials conjugate to the positive spin Teukolsky
perturbations $\{
\Psi^{(2)},\Psi^{(1)},\Psi^{(1/2)} \}$ 
but satisfy exactly the same master equation \eqref{MasterEq} as
$\lp -\Psi_{2}\rp^{-\frac{2s}{3}}\Psi^{(s)}$ for {\it negative}
spin. Similarly, $\Psi_{\rm H}^{(s)}=\{ \Psi_{\rm H}^{(2)},\Psi_{\rm
H}^{(1)},\Psi_{\rm H}^{(1/2)} \}$ are the Hertz potentials conjugate
to the negative spin Teukolsky perturbations
$\{\Psi^{(-2)},\Psi^{(-1)},\Psi^{(-1/2)} \}$
but the positive spin Hertz potential $\lp
-\Psi_{2}\rp^{-\frac{2s}{3}}\Psi_{\rm H}^{(s)}$ obeys the same
master equation as $\Psi^{(s)}$ for {\it positive} spin. In short,
the Hertz potential obeys the same master equation as its conjugated
Teukolsky field but with spin sign traded. This relation is better
clarified if we use Tables \ref{Table:Teuk} and \ref{Table:Teuk2}.
\begin{table}[th]
\begin{eqnarray}
\nonumber
\begin{array}{||c||||c|c||c|c||c|c||c||}\hline\hline\hline\hline
 \Psi_{\rm H}^{(s)}  &  \Psi_{\rm H_0}^{(-2)} & \lp -\Psi_{2}\rp^{-\frac{4}{3}} \Psi_{\rm H_4}^{(2)}
 & \Psi_{\rm H_0}^{(-1)} & \lp -\Psi_{2}\rp^{-\frac{2}{3}} \Psi_{\rm H_2}^{(1)}
 & \Psi_{\rm H_0}^{(-1/2)} & \lp -\Psi_{2}\rp^{-\frac{1}{3}} \Psi_{\rm H_1}^{(1/2)}  &  \Phi \\
\hline \hline
 s & -2 & 2 &  -1 & 1 & -\frac{1}{2} & \frac{1}{2} & 0 \\
\hline\hline\hline\hline
\end{array}
\end{eqnarray}
\caption{Spin $s$ Hertz fields $\Psi_{\rm H}^{(s)}$ that satisfy the
master equation \eqref{MasterEq} with no source term.}
\label{Table:Teuk2}
\end{table}

Assuming perturbations for the Hertz potentials of the form
\begin{equation}
\Psi_{\rm H}^{(s)}= \left\{
\begin{array}{ll}
e^{-i\omega t}e^{i m\phi}R_{lm\omega}(r)S_{lm}(\theta)\,, & \qquad s\leq 0 \,,\label{HertzAnsatz} \\
e^{-i\omega t}e^{i m\phi}R_{lm\omega}(r)S_{lm}(\theta) \lp
-\Psi_{2}\rp^{-\frac{2s}{3}} \,, & \qquad s\geq 0 \,,
\end{array}
\right.
\end{equation}
where $R_{lm\omega}(r)$ further satisfies \eqref{RadialAns},
equation \eqref{MasterEq} separates into an angular and radial
equations. The angular equation is \eqref{AngEq} with $C=m/2$, and
the radial equation is \eqref{RadialEq}. Its solution is given by
\eqref{RadialSol}.

As stated above, given the Hertz potential for the gravitational field
there is a unique map between it and the metric perturbations
\cite{Cohen:1975,Chrzanowski:1975,Cohen:1979,Wald}. A similar map
exists between the spin $s=\pm 1, \pm 1/2$ Hertz potentials
and the Maxwell and Weyl fermionic vector perturbations, but we
leave the discussion of these cases to Appendix \ref{sec:Wald}. In
the ingoing radiation gauge the metric perturbation in NP notation is given by (see Appendix
\ref{sec:Wald})
\begin{eqnarray}
h_{\mu\nu}^{\scriptstyle IRG}\!\!\!\!&=& \!\!\!\!{\biggl \{}
\ell_{(\mu}m_{\nu)} \left[
(D+3\epsilon+\overline{\epsilon}-\rho+\overline{\rho})
(\delta+4\beta+3\tau)
 +(\delta+3\beta-\overline{\alpha}-\tau-\overline{\pi})(D+4\epsilon+3\rho)
 \right]  \nonumber \\
 && -\ell_\mu \ell_\nu (\delta+3\beta+\overline{\alpha}-\overline{\tau})
(\delta+4\beta+3\tau) -m_\mu m_\nu
(D+3\epsilon-\overline{\epsilon}-\rho) (D+4\epsilon+3\rho) {\biggl
\}}\Psi_{\rm H}+{\rm c.c.}\,, \nonumber\\
 &&
 \label{huvHertz}
\end{eqnarray}
and a similar correspondence exists between the Hertz potential and
the metric perturbations $h_{\mu\nu}^{\scriptstyle ORG}$ in the
outgoing radiation gauge. (See the second relation of
\eqref{MapHertzHuv}.) One can check that \eqref{huvHertz} indeed satisfies the
linearized Einstein's equations for a traceless metric perturbation:
\eq -\nabla_\alpha\nabla^\alpha h_{\mu\nu}- 2R_{\mu\alpha\nu\beta}
h^{\alpha\beta}
 + 2g^{\alpha\beta}
\nabla_{(\mu}\nabla_{|\alpha |} h_{\nu ) \beta}=0\,.
  \label{Lichnerowicz}
\eeq

\subsection{Behaviour of solutions}

The basis vector fields  $\ell$ and $n$ are globally well-defined.
However, the vector field $m$ is singular at $\theta=0,\pi$.
Nevertheless, one can check that angular dependence of the Hertz
potential contains a sufficiently high power of $\sin \theta$ to
ensure that the above metric perturbation is smooth at
$\theta=0,\pi$.

The asymptotic behaviour of the Hertz potential $\Psi_{\rm H}^{(\pm
2)}$ can be obtained using \eqref{RadialAns} and \eqref{Phifalloff}.
Use of \eqref{huvHertz} yields then for the
asymptotic $h_{\mu\nu}$ behaviour (rows and columns follow the
order: $\{ t,r,\theta,\phi\}$)
 \eq \label{BChuv:IngoingGauge}
h_{\mu\nu}^{\scriptstyle IRG} \sim  r^{\frac{3}{2}\pm \frac{1}{2} \eta } \left(
  \begin{array}{ccccc}
 \mathcal{O}({1}) & \mathcal{O}({\frac{1}{r^2}})  &  \mathcal{O}({1\over r})  &
\mathcal{O}({1\over r})  \\
  \, &  \mathcal{O}({1\over r^4})& \mathcal{O}({1\over r^3}) &
\mathcal{O}({1\over r^3})  \\
 \, & \, & \mathcal{O}({1\over r^2})& \mathcal{O}({1\over r^2}) \\
 \, & \, & \,  &  \mathcal{O}({1\over r^2})  \\
  \end{array}
\right)\ , \eeq
where $\eta$ is given by \eqref{eqn:etadef}. Exactly the same result
is obtained  in the outgoing radiation gauge. In
\eqref{BChuv:IngoingGauge} we have not imposed any boundary
condition. These were discussed in subsection \ref{sec:BCs}; \eg for
$\eta^2>0$, the lower sign would correspond to normal modes.

We shall now compare the above asymptotic behaviour of metric
perturbations with  the GHSS fall-off conditions. The $tr$ and
$t\theta$ components are the most restrictive. For these to satisfy
the fall-off conditions, $\eta$ must be real, so traveling waves are
excluded, we must use normalizable boundary conditions (i.e. the
lower sign choice) and we need $\eta \ge 3$. Recall that there are
normal modes with $\eta=2.74$, so it appears that the GHSS fall-off
conditions exclude some of the normal modes.\footnote{It is
conceivable that a  gauge transformation could be used to bring a
mode violating the fall-off conditions to one that satisfies these
conditions but  this seems unlikely, especially for traveling
waves.}

As emphasized in the introduction, at the nonlinear level, we expect
that interactions will lead to modes corresponding to traveling
waves ($\eta^2<0$) being excited, which would lead to a violation of
the GHSS fall-off conditions. The only modes that escape this
conclusion are the axisymmetric ones (which have with $\eta =
2l+1$), which always obey the GHSS boundary conditions. Axisymmetric
modes form a consistent truncation of the full set of modes in the
sense that linearized axisymmetric modes will not excite
non-axisymmetric modes at next order in perturbation theory.

\subsection{The energy}

\subsubsection{Massless scalar field}

We want to compute the energy associated with the gravitational
perturbations that we found in the previous subsection. Since this
will involve a rather lengthy calculation, we shall start with the
conceptually simpler case of a massless complex scalar field:
 \eq \Box \Phi =0. \eeq
The canonical energy momentum tensor is given by
 \eq
T_{\mu\nu}=\nabla_{(\mu}\Phi\nabla_{\nu)}\Phi^\ast-\frac{1}{2}\bar{g}_{\mu\nu}\nabla_\alpha\Phi\nabla^\alpha\Phi^\ast.
\label{scalarTuv}
 \eeq
Let $\Sigma$ be a spacelike hypersurface with future-directed unit normal $n^\mu$.
Then, given any Killing vector $\xi^\mu$, we
can define the associated conserved charge
%
\eq Q_\xi[\Phi]=\int_{\Sigma}d^3x \sqrt{-\gamma} \,
T^{\mu\nu}n_{\mu}\xi_\nu, \label{def:charge}\eeq
where $\gamma_{\mu\nu}=\bar{g}_{\mu\nu}-n_\mu n_\nu$ is the induced
metric on $\Sigma$. We shall choose $\Sigma$ to be a surface of
constant $t$ in the NHEK geometry. The conserved charges of interest
are the energy $\mathcal{E}$, for $\xi=\partial/\partial t$, and the
angular momentum $\mathcal{J}$, for $\xi=-\partial/\partial \phi$.
(The latter is the $U(1)$ charge of GHSS.) Written out explicitly,
these are
\begin{eqnarray}
&& \hspace{-0.5cm}\mathcal{E} = M^2 \int_{0}^{2\pi}d\phi
\int_{0}^{\pi}d\theta
\int_{-\infty}^{\infty}dr\left[\frac{\left|\partial_t\Phi
\right|^2}{1+r^2}+\left(1+r^2\right)\left|\partial_r\Phi
\right|^2+\left|\partial_{\theta}\Phi \right|^2+\left(
\frac{1}{\Lambda(\theta)^2}-\frac{r^2}{1+r^2}\right)\left|\partial_{\phi}\Phi
\right|^2\right],\nonumber \\
 &&\hspace{-0.5cm} \mathcal{J} =M^2\int_{0}^{2\pi}d\phi \int_{0}^{\pi}d\theta
\int_{-\infty}^{\infty}dr \,\frac{1 }{1+r^2} \left[
2r\left|\partial_{\phi}\Phi \right|^2-\left(\partial_t\Phi
\partial_{\phi}\Phi^*+\partial_t\Phi^*
\partial_{\phi}\Phi\right)\right].
\end{eqnarray}
In the energy integrand, all the terms except the last are
manifestly positive. The last is proportional to
$|\partial_t|^2=g_{tt}$ and thus is positive only outside the
ergosphere where $1+r^2-r^2\Lambda(\theta)^2>0$. The energy can thus
be negative (for a rigorous proof of this, see Ref. \cite{friedman}).

Consider a general superposition of normalizable modes (recall  that
$n$ is defined by the frequency quantization \eqref{FreqSpectrum}):
\eq \label{eqn:super}
 \Phi(x) = \sum_{nlm} \left( a_{nlm} \Phi_{nlm}(x) + b_{nlm} \Phi_{nlm}(x)^*
 \right).
\eeq First, we shall show that the conserved charges associated with
such a solution can be decomposed into a sum of conserved charges of
the individual modes.

A charge integral $Q_\xi[\Phi]$ can be regarded as defining a
(typically indefinite) norm on the space of solutions of the wave
equation. Given a norm $|{}|$, it can be ``polarized" to obtain a
Hermitian scalar product $({},{})$: the real and imaginary parts of
$(u,v)$ are given by $(|u+v|-|u|-|v|)/2$ and $(|iu+v|-|u|-|v|)/2$
respectively. In our case, polarizing the charge integral defines a
scalar product $(\Phi_1,\Phi_2)_\xi$, antilinear in $\Phi_1$ and
linear in $\Phi_2$. Since the norm is conserved, so will be the
scalar product. Note that $(\Phi,\Phi)_\xi = Q_\xi[\Phi]$.

We shall now argue that modes with different $(nlm)$ are orthogonal with respect to this scalar product.
The scalar product has the form
\eq
 (\Phi_1,\Phi_2)_\xi = \int_\Sigma d^3 x Q^{\mu\nu}(x) \partial_\mu \Phi_1^* \partial_\nu \Phi_2,
\eeq where $Q^{\mu\nu}$ is preserved by any Killing vector field
that  commutes with $\xi$. Now let $\eta$ be such a Killing field.
We can then write \eq
 (\Phi_1, -i {\cal L}_\eta \Phi_2)_\xi - (-i {\cal L}_\eta \phi_1, \phi_2)_\xi
 = \int_{\Sigma} {\cal L}_\eta ( Q^{\mu\nu}(x) \partial_\mu \Phi_1^* \partial_\nu
 \Phi_2).
\eeq If the RHS vanishes then this shows that $-i {\cal L}_\eta$ is
self-adjoint  with respect to the this scalar product. For NHEK, we
take $\xi=\partial/\partial t$ or $\partial/\partial \phi$. Taking
$\eta=\partial/\partial t$, the RHS vanishes because the scalar
product is conserved, and hence independent of $t$. Taking
$\eta=\partial/\partial \phi$, the RHS vanishes because it is a
total derivative on $\Sigma$. It follows that modes with different
$\omega$ or diffferent $m$ will be orthogonal with respect to this
scalar product. Hence, in calculating the charge associated with
\eqref{eqn:super}, there are no cross-terms in the charge arising
from modes with different $\omega$ or $m$ (in particular, there are
no cross-terms between the positive and negative frequency parts of
\eqref{eqn:super}).

Now consider the $l$-dependence. Since $l$ is not associated with a
Killing symmetry of the background, we cannot use the above
argument. Instead,  for separable solutions, the angular dependence
will be given by \eqref{AngEq} with $s=0$. This equation is
self-adjoint, so  two solutions with different values of
$\Lambda_{lm}^{(s)}$ will be orthogonal with respect to the measure
$\sin\theta$, i.e.,
 \eq \int_0^\pi \dd\theta
\sin\theta \,S_{l_1 m}^{(s)}(\theta)S_{l_2 m}^{(s)}(\theta)^*
\propto \delta_{l_1 l_2}. \eeq Fortunately, it turns out that $\sin
\theta$ is precisely the measure  that arises in the scalar products
associated with the energy and angular momentum.

From these results, we see that no cross-terms between modes with
different $(nlm)$ contribute to the energy and angular momentum.
Substituting \eqref{eqn:super} into \eqref{def:charge} for
$\xi=\partial_t$ gives the energy as a sum over contributions from
individual modes: \eq \mathcal{E}  = \sum_{n l m} \mathcal{E}_{n l
m} (|a_{nlm}|^2+|b_{nlm}|^2), \eeq where \eq \mathcal{E}_{n l m}
 \equiv 4\pi M^2 \omega_{n l m}
  \int_0^\pi  \dd\theta \sin\theta |S_{lm}^{(0)}(\theta)|^2
    \int_{-\infty}^{+\infty}\dd r |R_{n l m}^{(0)}(r)|^2\frac{\omega_{n l m}+ m
    r}{1+r^2}\,.
\eeq Note that $\mathcal{E}_{nlm}$ is manifestly positive only when
$m=0$.  However, we have evaluated the radial integral above for
many cases, namely for $0\leq l\leq 10$, $-l\leq m\leq l$ and
$0\leq n\leq 10$. In all these cases, it is positive. Hence, for a
massless complex scalar field in the NHEK geometry, the energy of an
arbitrary superposition of normalizable modes is positive.

The angular momentum can be similarly decomposed:
\eq
\mathcal{J}  = \sum_{n l m} \mathcal{J}_{n l m}  (|a_{nlm}|^2-|b_{nlm}|^2),
\eeq
where we find the simple result
\eq
  \frac{\mathcal{J}_{n l m}}{\mathcal{E}_{n l m}} = \frac{m}{\omega_{nlm}}.
\eeq

\subsubsection{Gravitational perturbations}

The energy of gravitational perturbations is calculated from the
Landau-Lifshitz ``pseudotensor" defined as follows. Consider metric perturbations
$h_{\mu\nu}$ around NHEK up to second order in the amplitude, \eq
 g_{\mu\nu}=\bar{g}_{\mu\nu}+h_{\mu\nu}=\bar{g}_{\mu\nu}+h^{(1)}_{\mu\nu}+h^{(2)}_{\mu\nu}+\mathcal{O}{(h^3)},
\eeq The linearized Einstein equation is\footnote{Written out, this
takes  the standard Lichnerowicz form. If we assume that $h^{(1)}$
is traceless then this equation reduces to \eqref{Lichnerowicz}.}
 \eq
G^{(1)}_{\mu\nu}[h^{(1)}]=0 \,. \label{Einstein1st}
 \eeq
At second order, the Einstein equation relates terms linear in
$h^{(2)}$  to terms quadratic in $h^{(1)}$:
 \eq
G^{(1)}_{\mu\nu}[h^{(2)}] = -G^{(2)}_{\mu\nu}[h^{(1)}]\ \equiv 8\pi G T_{\mu\nu}[h^{(1)}] \,, \label{Einstein2nd}
 \eeq
where the RHS is quadratic in $h^{(1)}$. Written out explicitly, for
traceless perturbations it reads (here, we use the notation
$h_{\mu\nu}\equiv h_{\mu\nu}^{(1)}$)
 \begin{eqnarray}
 8 \pi G T_{\mu\nu}&=& -\frac{1}{2}{\biggl[}\frac{1}{2}\lp \nabla_\mu h_{\alpha\beta}\rp \nabla_\nu
 h^{\alpha\beta}
 + h^{\alpha\beta}\lp \nabla_\nu \nabla_\mu h_{\alpha\beta}+\nabla_\alpha \nabla_\beta h_{\mu\nu}
   - \nabla_\alpha \nabla_\mu h_{\nu\beta} -\nabla_\alpha \nabla_\nu
   h_{\mu\beta}\rp \nonumber\\
&& \hspace{0.7cm}+  \nabla_\alpha h^{\beta}_{\:\:\mu}\lp
\nabla^\alpha h_{\beta\nu}-\nabla_\beta h^{\alpha}_{\:\:\nu} \rp -
\nabla_\alpha h^{\alpha\beta}\lp \nabla_\mu h_{\beta\nu}+\nabla_\nu
h_{\mu\beta}-\nabla_\beta h_{\mu\nu} \rp{\biggl]}\nonumber\\
&& +\frac{1}{4}\,\bar{g}_{\mu\nu}{\biggl[}\frac{1}{2}\lp
\nabla_\gamma h_{\alpha\beta}\rp \nabla^\gamma
 h^{\alpha\beta}
 + h^{\alpha\beta}\lp \nabla_\gamma \nabla^\gamma h_{\alpha\beta}
   - 2 \nabla_\alpha \nabla^\gamma h_{\gamma\beta} \rp \nonumber\\
&&\hspace{1.3cm} +  \nabla_\alpha h^{\beta\gamma}\lp \nabla^\alpha
h_{\beta\gamma}-\nabla_\beta h^{\alpha}_{\:\:\gamma} \rp - 2 \lp
\nabla_\alpha h^{\alpha\beta}\rp \nabla^\gamma
h_{\beta\gamma}{\biggl]}. \label{gravT}
\end{eqnarray}
 We now define the conserved charges $Q_\xi[h^{(1)}]$ associated
with the  first order perturbation exactly as in \eqref{def:charge}
with $\xi=\partial/\partial t$, $-\partial/\partial\phi$ giving the
energy and angular momentum respectively.

Recall that $h^{(1)}$ is related to the Hertz potential by equation
\eqref{huvHertz}, which is second order in derivatives. It follows
that the conserved charges are given by integrals of quantities that
are sixth order in derivatives. Hence calculating these charges
involves very lengthy calculations, which we have performed using
computer algebra.

We consider a Hertz potential corresponding to an arbitrary
superposition of normal modes:
\eq
 \Psi_H^{(s)}(x) = \sum_{nlm}  \left( a_{nlm} \Psi_{nlm}^{(s)+} (x) + b_{nlm} \Psi_{nlm}^{(s)-} (x) \right),
\eeq
where $s=\pm 2$ and the superscript $\pm$ refers to positive and negative frequency respectively.

As in the scalar field, case the conserved charges can be used to
define a  scalar product $(,)_\xi$ between solutions of the
linearized Einstein equation. The only significant difference here
is that the metric perturbation is real, so the scalar product will
also be real.\footnote{The polarization formula is
$(u,v)=(1/2)(|u+v|-|u|-|v|)$. We could have chosen to work with
complex  modes $h_{\mu\nu}$, for which the negative frequency modes
are complex conjugates of the positive frequency modes. However,
then we would have had to take account of two different
polarizations for the gravitational modes.} One can argue exactly as
in the scalar field case that ${\cal L}_\eta$ is anti-self-adjoint
with respect to this scalar product if $\eta$ is a Killing field
that commutes with $\xi$: $(h,{\cal L}_\eta k)=(-{\cal L}_\eta
k,h)$. It follows that the operator $-{\cal L}_\eta^2$ is
self-adjoint and hence linearized metric perturbations with
different $\omega^2$ or different $m^2$ must be orthogonal.

Given the complexity of the charge integrals, we have not succeeded
in  demonstrating that modes with different $l$ are orthogonal in
the same way that we did for the scalar field. However, note that
$\omega$ depends on $l$ in a very complicated way (through the
eigenvalues $\Lambda^{(2)}_{lm}$ which must be found numerically).
Hence it seems very unlikely that modes with different $l$ could
have the same $\omega^2$. Therefore the orthogonality of modes with
different $\omega^2$ should ensure the orthogonality of modes with
different $l$. An exception are the axisymmetric ($m=0$) modes,
which have $\omega = n+l+1$ so modes with the same $n+l$ have the
same $\omega$. However, the axisymmetric modes are the ``least
dangerous" as far as the possibility of negative energy is concerned
so we shall not worry about this further, and simply assume that all
modes with different $l$ will be orthogonal.

We now turn to our calculation of the conserved charges associated
with individual modes. These charges are most easily computed in NP
tetrad, since this is the basis in which the metric perturbation
takes the relatively simple form  \eqref{huvHertz}, although the
explicit expressions for the components are still too long to be
written here. Using Mathematica, the separated equation of motion
can be used to reduce the integrands of the charge integrals to
expressions first order in derivatives, which were then calculated
numerically using Mathematica's NIntegrate function.

We have calculated the energy of normal modes with $l=2,3,4,5,6$ for
all  allowed values of $m$, $n=0,1,2,3,4,5,6$ and both positive and
negative frequency. In all cases it comes out positive. This is the
main result of this section.

\begin{figure}[t]
\centerline{\includegraphics[width=.30\textwidth]{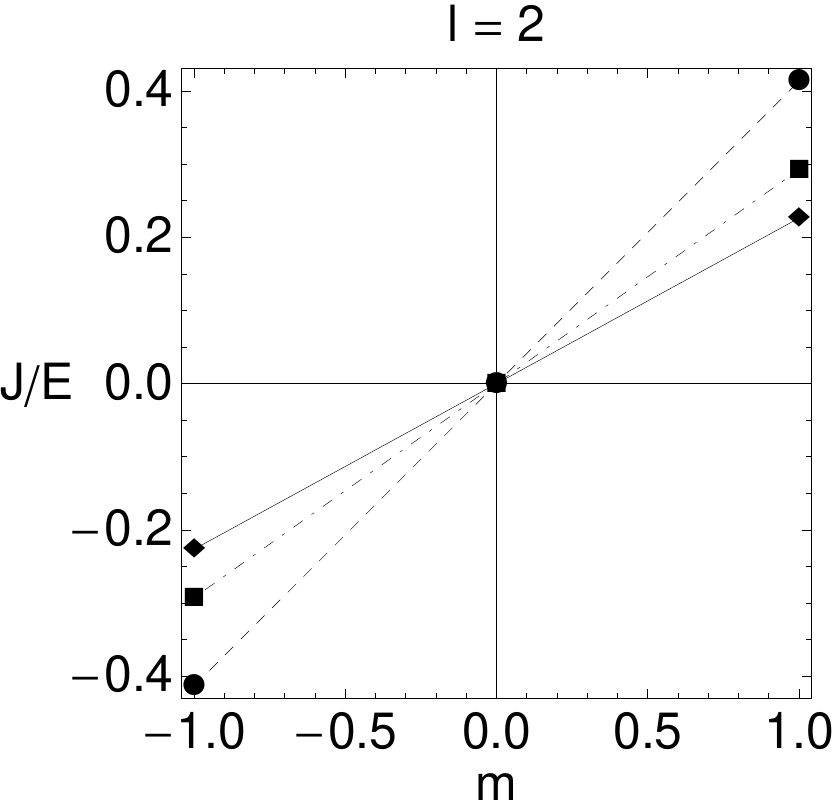}
\hspace{2cm}\includegraphics[width=.30\textwidth]{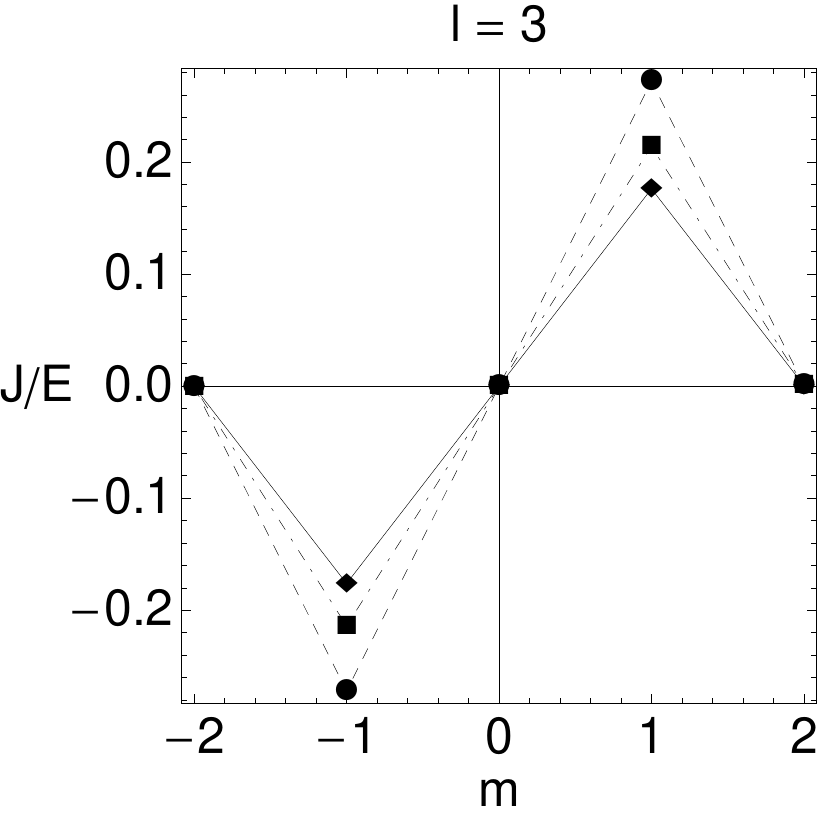}}
\caption{Ratio of the conserved charges $\mathcal{J}/\mathcal{E}$
for spin $|s|=2$ perturbations as a function of the azimuthal
angular number $m$ for  a) $l=2$ and b) $l=3$.  We only consider
normal modes, \ie values of $m$ that yield $\eta^2>0$ as defined in
\eqref{eqn:etadef}. Data points corresponding to $n=0,1,2$ are
plotted, with the solid line representing $n=2$ and the dashed line
$n=0$.} \label{fig:ratioJE}
\end{figure}

\begin{figure}[t]
\centerline{\includegraphics[width=.30\textwidth]{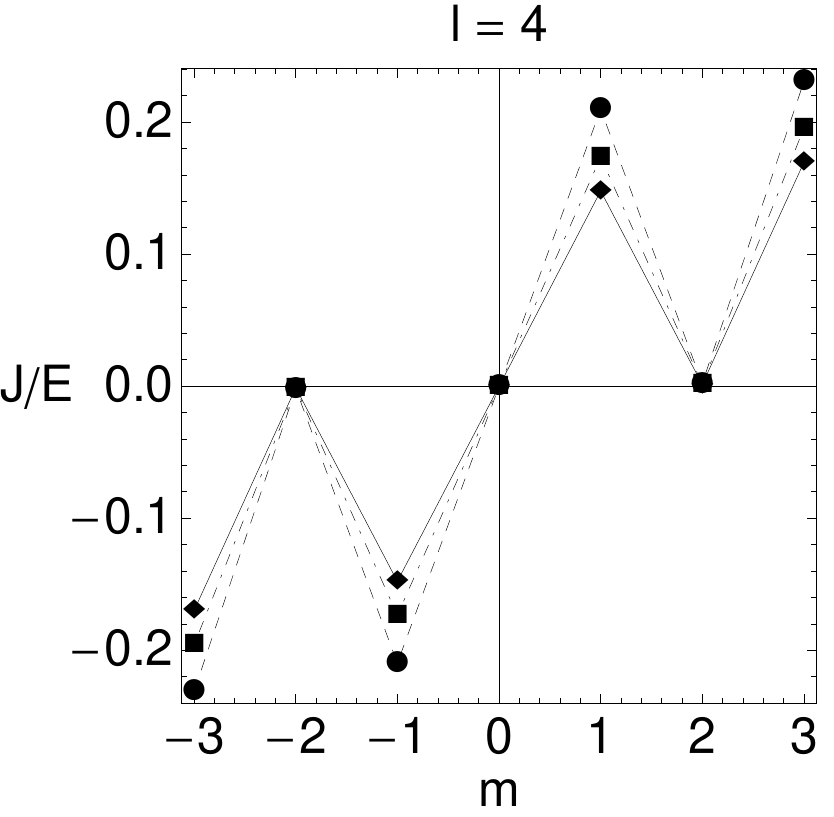}
\hspace{2cm}\includegraphics[width=.30\textwidth]{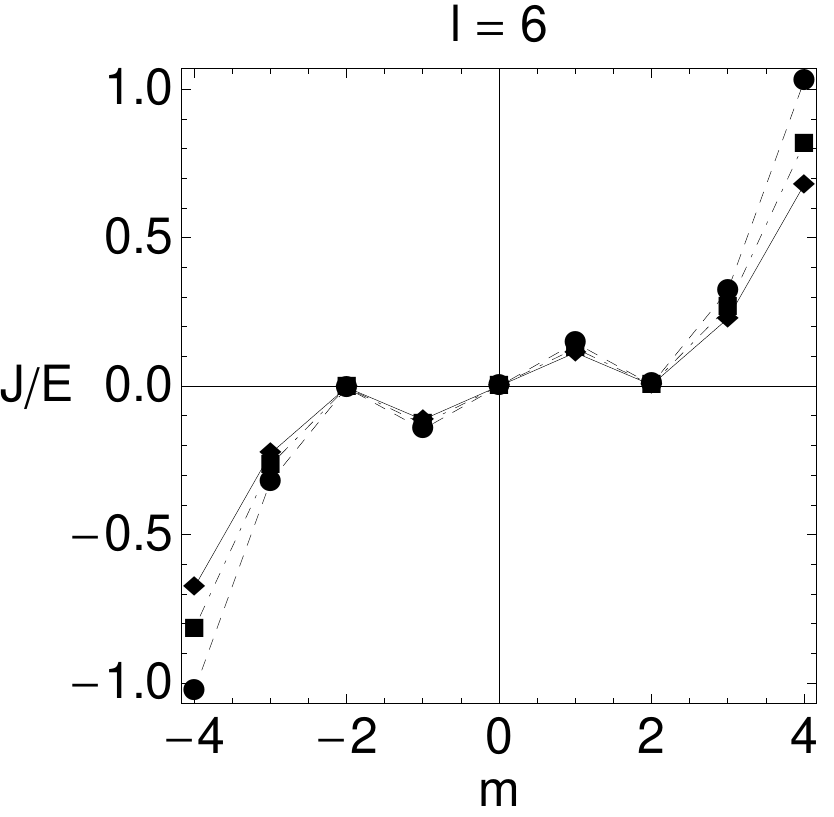}}
\caption{Ratio of the conserved charges $\mathcal{J}/\mathcal{E}$
for spin $|s|=2$ perturbations as a function of the azimuthal
angular number $m$ for  a) $l=4$ and b) $l=6$. }
\label{fig:ratioJE2}
\end{figure}

The only difference between positive and negative frequency modes is
the sign of $\mathcal{J}$ so we focus on the positive frequency
case. The numerical value of the energy depends on the normalization
of the Hertz potential. However, the ratio of the conserved charges,
$\mathcal{J}_{nlm}/\mathcal{E}_{nlm}$ is normalization independent.
This ratio as a function of $m$ and $n$ for fixed $l$ is displayed
in Figs. \eqref{fig:ratioJE}-\eqref{fig:ratioJE2}. Note that
$\mathcal{J}_{nl(-m)}=-\mathcal{J}_{nlm}$ and that
$\mathcal{E}_{nl(-m)}=\mathcal{E}_{nlm}$.
Moreover, for any given $l$ the ratio $|\mathcal{J}|/\mathcal{E}$
always has a (non-vanishing) minimum at $|m|=|s|=2$ (the ratio
decreases with $n$ but only slowly, so this is not apparent in the
plots). The corresponding modes also exhibit special behaviour in
the Kerr geometry: Teukolsky and Press \cite{Teukolsky:1974yv} found
that for a given black hole rotation and wave frequency, the modes
whose energy is most absorbed or superradiantly amplified are
precisely those with $l=|m|=|s|$.

\section{Discussion: second order perturbations}
\label{sec:discussion}

We now turn to the question of what happens if we go beyond first
order in perturbation theory. The second order metric perturbation
$h^{(2)}$ is determined by solving \eqref{Einstein2nd}.\footnote{If
we considered scalar field perturbations then the gravitational
backreaction of the perturbation would be governed by the same
equation with $T_{\mu\nu}$ the scalar field stress tensor.} We are
not going to attempt to solve this equation. Instead, following
recent work on TMG \cite{Maloney:2009ck} we consider the conserved
charges.

So far, we have worked with conserved charges defined via bulk integrals quadratic in $h^{(1)}$. However, one can also define conserved charges via boundary integrals, indeed these are the charges discussed by GHSS. So first we shall explain how they are related to our bulk integrals. Consider a 1-parameter family of exact vacuum solutions $g(\lambda)$, where $g(0) \equiv \bar{g}$ is the NHEK metric. Let $h^{(1)}=g'(0)$ and $h^{(2} = (1/2) g''(0)$. ($h^{(1)}$ is the linearized solution arising from the linearization of $g(\lambda)$, $h^{(2)}$ is the second order correction.) Owing to the unusual fall-off conditions, the conserved charge $Q_\xi(\lambda) \equiv Q_\xi [g(\lambda)]$ associated to a generator $\xi$ of the asymptotic symmetry group are defined by integrating the following expression:
\eq
\label{dQ}
 \frac{dQ_\xi}{d\lambda} = {\cal Q}_\xi[g'(\lambda),g(\lambda)]
\eeq
where
\begin{eqnarray}
  {\cal Q}_\xi[h,g] &\equiv& -{1\over 32 \pi
G}\int_{\partial \Sigma} \epsilon_{\alpha\beta\mu\nu} \big[ \xi^\nu
\nabla^\mu h - \xi^\nu \nabla_\sigma h^{\mu\sigma}
 + \xi_\sigma \nabla^\nu h^{\mu\sigma} + {1\over 2}\,h
\nabla^\nu \xi^\mu
 - h^{\nu\sigma} \nabla_\sigma \xi^\mu  \nonumber \\
& & \hspace{3.2 cm} + {1\over 2} h^{\sigma\nu}(\nabla^\mu\xi_\sigma
+ \nabla_\sigma\xi^\mu)\big] dx^\alpha\wedge dx^\beta \,.
\label{chargeXi}
\end{eqnarray}
Now, our gravitational normal modes decay sufficiently fast that they give ${\cal Q}_\xi[h^{(1)},\bar{g}]=0$, hence $dQ_\xi/d\lambda=0$ at $\lambda=0$. This is no surprise since we know that the energy should be quadratic in $h^{(1)}$. Hence we have to go to next order, and calculate $(1/2)d^2 Q_\xi/d\lambda^2$ at $\lambda=0$. This can be done by differentiating \eqref{dQ}, which gives a sum of a part linear in $h^{(2)}$, equal to ${\cal Q}_\xi[h^{(2)},\bar{g}]$, and a part quadratic in $h^{(1)}$. However, the normal modes decay so fast that this second part vanishes. Hence, to second order in $\lambda$, we have that
\eq
 Q_\xi(\lambda) = \lambda^2 {\cal Q}_\xi [h^{(2)},\bar{g}].
\eeq
Now, assuming that $\xi$ is a Killing field of the background, a standard manipulation \cite{Abbott:1981ff,Barnich:2001jy,Deser:2002jk} based on the second-order Einstein equation \eqref{Einstein2nd} enables one to rewrite this surface integral as the bulk integral quadratic in $h^{(1)}$ that we used in the previous section.

One subtlety is that the NHEK geometry
has two boundaries (at $r = \pm \infty$). The bulk
integral for the charge will be the sum of the two surface
integrals. Hence, if the first order perturbation gives a non-zero
conserved charge, then the second order perturbation $h^{(2)}$ must
decay sufficiently slowly for these surface integrals to be
non-zero.

Consider initial data (say at $t=0$) for a first order perturbation $h^{(1)}$ that
is  of compact support (this will necessarily involve harmonics
$(l,m)$ corresponding to traveling waves). How will the second order
perturbation sourced by this first order perturbation behave? Near
infinity (at least at early times), $h^{(2)}$ must satisfy the
source free linearized Einstein equation, i.e., the same equation as
the first order perturbation. Hence the behaviour of $h^{(2)}$ near
infinity should be the same as that of a first order perturbation.
However, none of the first order normal modes decays sufficiently
slowly to make a non-vanishing contribution to the surface integrals
for the charges, e.g., a non-vanishing contribution to the energy
would require $\eta \le 1$ in \eqref{BChuv:IngoingGauge} (with the
lower sign choice) whereas we have seen that normal modes have $\eta
> 2.74$. A non-vanishing contribution to the angular momentum
requires $\eta \le -1$. Therefore $h^{(2)}$ does not behave like a
normal mode at infinity. Furthermore, even the traveling waves decay
too slowly to contribute to the surface integral for the angular
momentum. So what linearized solution does  $h^{(2)}$ behave like
near infinity?

Precisely the same issue arises for a Kerr black hole. Gravitational
perturbations with $l \ge 2$ decay too fast to contribute to the
surface integrals for the energy or angular momentum. For Kerr, the
resolution is that the Teukolsky or Hertz potential formalisms miss
certain modes, specifically those modes that preserve the type D
condition to first order. For Kerr, it has been shown that the only
such perturbations correspond to deformations towards a nearby type
D solution \cite{wald}. The nearby solutions are: the Kerr solution
with different $(M,J)$, the Kerr-NUT solution, and the spinning
C-metric. The latter perturbations are excluded by asymptotic
boundary conditions or regularity. Hence, for Kerr, one must add by
hand the non-dynamical modes corresponding to infinitesimal
variations in the mass and angular momentum of the black hole, which
we can regard as $l=0$ and $l=1$ perturbations respectively. Clearly
these will decay at an appropriate rate to contribute to the surface
integrals.

This suggests that, in our case, the fall-off of $h^{(2)}$ will be
the same as that of linearized modes that preserve the type D
property. There are two classes of such modes:  (i) modes that are
locally  gauge, i.e., locally of the form $\nabla_{(\mu} \eta_{\nu)}$, and
(ii) modes corresponding to a non-trivial deformation towards a type
D solution continuously connected to NHEK.

Consider first the case that $h^{(2)}$ behaves asymptotically as a linearized
mode that is locally gauge. By this we mean that, in a
neighbourhood of the $S^2$ on which a boundary integral is computed,
$h^{(2)}$ is locally, but not globally, of the form $\nabla_{(\mu}
\eta_{\nu)}$.\footnote{If it were globally a gauge transformation
then it would give a vanishing boundary integral since, for a Killing field $\xi$, ${\cal Q}_\xi[h,\bar{g}]$ is invariant under $h_{\mu\nu} \rightarrow h_{\mu\nu}+\nabla_{(\mu} \eta_{\nu)}$
(even if $\eta$ is a non-trivial element of the asymptotic symmetry group).} This is
precisely what happens for Einstein gravity in $AdS_3$, for example,
where $\eta_{\mu}$ cannot be globally defined on the $S^1$ boundary.
Could the same thing happen here? One might consider infinitesimal
diffeomorphisms of the GHSS form $\epsilon(\phi) \partial/\partial
\phi - \epsilon'(\phi) r \partial/\partial r$ and, instead of taking
$\epsilon$ to be periodic in $\phi$ (which would be globally
defined), take $\epsilon(\phi)=\phi$, which leads to a metric
perturbation independent of $t$ and $\phi$. However, this has the
effect of introducing a conical singularity into the metric near
infinity (at the poles of the $S^2$), which does not seem
appropriate.

This ``locally gauge" behaviour would arise from solutions that are
obtained by identifications of the NHEK background (in the same way
that the BTZ black hole is obtained as an identification of
$AdS_3$). Could one obtain a ``NHEK black hole" by identifying the
NHEK geometry in some way? Assuming any such identification acts
only on the surfaces of constant $\theta$, the possibilities have
been well-studied \cite{Anninos:2008fx}, and there appears to be no
candidate free of pathologies such as conical singularities or
closed timelike curves.

Consider then, the second possibility, that $h^{(2)}$ behaves asymptotically as a linearized mode corresponding to a deformation towards a nearby type D solution. What solutions are
there? Using Kinnersley's classification of type D solutions
\cite{Kinnersley:1969zza}, the only such solutions appear to be:
NHEK with a change in the angular momentum, the full (asymptotically
flat) Kerr solution, or the near-horizon geometry of the extremal
spinning C-metric. The latter has a conical singularity and so
presumably must be excluded.

It appears that the only candidate for a ``$l=1$" mode, i.e., a mode
contributing to the surface integral for angular momentum, is the
perturbation that corresponds to a change in the angular momentum of
the NHEK geometry ($J \rightarrow J + \delta J$ in \eqref{NHkerr}).
This violates the GHSS fall-off conditions. Hence it would appear
that, at second order, any perturbation with non-vanishing angular
momentum is excluded by the fall-off conditions.

What about the energy?  One can attempt to obtain a solution with
non-zero energy by taking a decoupling limit of the near-extremal
Kerr solution at fixed temperature and angular momentum. An
analogous decoupling limit of Reissner-Nordstr$\ddot{o}$m was
discussed in Ref. \cite{Maldacena:1998uz}. However, in the latter
case, it was shown that, even with non-zero temperature, the
decoupling geometry is simply $AdS_2 \times S^2$. We find that the
same is true for Kerr: in Appendix \ref{app:decouple}, we show that
the decoupling limit at fixed non-zero temperature leads back to the
NHEK geometry. The explanation is presumably the same as in Ref.
\cite{Maldacena:1998uz}, namely that the extreme Kerr black hole has
a mass gap.

It appears that the only regular modes with non-zero energy
correspond to going to next order in the decoupling limit. This is
probably equivalent (up to a $SL(2,R)$ transformation) to retaining
the next to leading order term in the near-horizon limit leading
from extreme Kerr to NHEK. This clearly gives a solution $k^{(1)}$ of the
linearized Einstein equation. However, it violates the GHSS fall-off conditions, indeed at the fully nonlinear level it amounts to considering an asymptotically flat black hole rather than its
near-horizon limit.

If correct, this implies that, if the first order perturbation has
any non-zero energy (whether positive or negative) or angular
momentum then at second order there will be a violation of the GHSS
fall-off conditions.\footnote{
If one wanted to impose these fall-off conditions only at $r \rightarrow \infty$ but didn't care what happened at $r \rightarrow -\infty$ then one could always add to $h^{(2)}$ an appropriate multiple of $k^{(1)}$ to arrange this, because $k^{(1)}$ satisfies the linearized Einstein equation, and one is free to add to $h^{(2)}$ (which satisfies the inhomogeneous equation \eqref{Einstein2nd}) any solution of the linearized Einstein equation. Note that the boundary integrals associated with any solution of the linearized Einstein equation, e.g. $k^{(1)}$, must sum to zero.}
This would be a satisfying conclusion: one does
not have to worry about negative energy initial data, and the
positive energy condition is redundant (at least in perturbation
theory). What about initial data for a linearized gravitational
field with {\it vanishing} energy and angular momentum? There certainly exists initial data with this property. We have seen that
the normal modes have positive energy, so this data must involve
traveling waves. With outgoing boundary conditions, the linearized
theory predicts that these will disperse, leaving behind only normal
modes, with positive energy. If this extends to the nonlinear theory
then there still would be a problem since the final state would have
to violate the fall-off conditions. It seems to us that the only
solution is that, even though this initial data has vanishing
energy, the two boundary integrals for the energy would be non-zero, but opposite in sign. Hence one would still obtain  $h^{(2)}$ with the asymptotic
behaviour just discussed, and thereby violate the fall-off
conditions.

This reasoning suggests that, at the nonlinear
level,  there are no non-trivial (i.e. non-isometric to NHEK) solutions of
the Einstein equation that are continuously connected to NHEK, and
satisfy the GHSS fall-off conditions (see Ref. \cite{Maldacena:1998uz}
 for a proof of a similar result for $AdS_2 \times S^2$). This may imply that the only
solutions that satisfy the latter are related to NHEK by large gauge
transformations. However, in this case, the dual CFT would consist
purely of conformal descendents of the vacuum, which leads to a
problem with modular invariance. Alternatively, there might be
further solutions that are asymptotic to NHEK in the GHSS sense, but
not continuously connected to it. If so, it would be interesting to
find these solutions.



\section*{Acknowledgments}

We are grateful to Geoffrey Comp\`ere, Mihalis Dafermos, Edvin
Deadman,  Stefan Hollands, John Stewart, and especially Monica Guica
for discussions. OJCD acknowledges financial support provided by the
European Community through the Intra-European Marie Curie contract
PIEF-GA-2008-220197. HSR is a Royal Society University Research
Fellow. JES acknowledges support from the Funda\c c\~ao para a
Ci\^encia e Tecnologia  (FCT, Portugal) through the grant
SFRH/BD/22058/2005. This work was partially funded by FCT-Portugal
through projects PTDC/FIS/64175/2006 and CERN/FP/83508/2008.

\appendix

\section*{Appendices}

\setcounter{equation}{0}
 \section{Shear-free null geodesics and NP tetrad for NHEK}
 \label{sec:geodesic}

\subsection{NP quantities for NHEK in global coordinates  \label{sec:NPglobal}}

NHEK is a Petrov type D geometry and its NP null tetrad is found
by looking into the congruence of shear-free null geodesics
\cite{ChandrasekharBook}.

Geodesics are the paths that minimise the action associated with the
Lagrangian
\eq
 \mathcal{L}=\frac{1}{2}g_{\mu\nu}\,\frac{dx^\mu}{d\lambda}\frac{dx^\nu}{d\lambda}=\frac{\delta}{2}\,,
  \label{Lgeod}
\eeq
where $\lambda$ is an affine parameter, and $\delta=0,1$,
respectively, for null and time-like geodesics. Since the NHEK
geometry \eqref{NHkerr} is stationary, the energy $E$ and angular
momentum $L$ of the particle,
\eq
 p_t= g_{tt} \dot{t}+g_{t\varphi} \dot{\varphi} \equiv E \,, \qquad
  p_\varphi= g_{t\varphi} \dot{t}+g_{\varphi\varphi} \dot{\varphi} \equiv L\,,
  \label{EnAngMomDef}
\eeq
are conserved in a geodesic motion,  where $p_\mu\equiv \frac{d
\mathcal{L}}{dx^\mu}$ is the conjugated momentum. Equation
\eqref{EnAngMomDef} yields
\eq
 \dot{t}=\frac{E-Lr}{M^2 (1+\cos^2\theta)(1+r^2)} \,,\qquad
 \dot{\varphi}= -\frac{r(E-Lr)}{M^2
(1+\cos^2\theta)(1+r^2)}-\frac{L}{4M^2}\,\frac{
1+\cos^2\theta}{\sin^2\theta}
 \,.
  \label{EnAngMomCons}
\eeq

The Hamilton-Jacobi equation for the geodesic motion on a geometry
$g_{\mu\nu}$ reads
\eq
 \frac{\partial S}{\partial \lambda}= H\lp x^\mu, \frac{\partial S}{\partial
x^\mu},\lambda
 \rp \,, \qquad
 H\lp x^\mu, \frac{\partial S}{\partial x^\mu},\lambda
 \rp =\frac{1}{2} g_{\mu\nu}\,\frac{\partial S}{\partial x^\mu}\frac{\partial
S}{\partial
 x^\nu}\,.
  \label{HamJac}
\eeq
Assuming a separation {\it ansatz} of the form
\eq
 S=\frac{1}{2}\,\delta\lambda +Et+L\varphi +S_r(r)+S_\theta(\theta)\,,
  \label{Sansatz}
\eeq
equation \eqref{HamJac} for the NHEK boils down to
\begin{eqnarray}
  &&  \lp \frac{\partial S_\theta}{\partial \theta} \rp^2=
\Theta(\theta)\,, \qquad \Theta(\theta)\equiv \Lambda
-L^2\,\frac{\lp 1+\cos^2\theta \rp^2}{4\sin^2\theta}-M^2\delta
\cos^2\theta\,;
\nonumber \\
  && (1+r^2)^2 \lp \frac{\partial S_r}{\partial r} \rp^2=
\mathcal{R}(r)\,, \qquad \mathcal{R}(r)\equiv \lp
E-rL\rp^2-(\Lambda+ M^2\delta)(1+r^2)\,;
 \label{HamJacEqs}
\end{eqnarray}
where $\Lambda$ is the separation constant. Typically one has
$\dot{r}^2\propto \mathcal{R}(r)$ and $\dot{\theta}^2  \propto
\Theta(\theta)$ and the conservation equations \eqref{EnAngMomCons}
give the remaining equations for $\dot{t}$ and $\dot{\varphi}$.

The shear-free principal null geodesics are found by requiring
$\dot{\theta} \propto \Theta(\theta)=0$ for $\delta=0$, which in our
case requires
\eq
 \Lambda= L^2\,\frac{\lp 1+\cos^2\theta_0 \rp^2}{4\sin^2\theta_0}\,,
  \label{HJcondition1}
\eeq
for a constant $\theta=\theta_0$. Moreover, these geodesics must
also keep $\dot{r}$, \ie
\eq
 \mathcal{R}= \lp
E-rL\rp^2-(1+r^2)L^2\,\frac{\lp 1+\cos^2\theta_0
\rp^2}{4\sin^2\theta_0}
  \label{HJcondition2}
\eeq
constant along the motion. Clearly this is possible for any $r$ and
$\theta_0$ only if $L=0$. The energy and angular momentum
conservation equations \eqref{EnAngMomCons} then require
\eq
 M^2 (1+\cos^2\theta_0) \dot{t}=\frac{E}{1+r^2} \,,\qquad
 M^2 (1+\cos^2\theta_0) \dot{\varphi}= -\frac{r E}{1+r^2}\,.
  \label{EnAngMomCons2}
\eeq
This Hamilton-Jacobi analysis then concludes that shear-free null
geodesics have the tangent vectors
\begin{eqnarray}
  && \ell^\mu \partial_\mu= \frac{1}{1+r^2}\,\partial_t+\partial_r -
\frac{r}{1+r^2}\,\partial_\varphi  \,, \nonumber \\
  && n^\mu \partial_\mu= \frac{1}{4M^2\Omega^2(\theta)}\lp
\partial_t-(1+r^2)\partial_r - r\partial_\varphi \rp \,,
 \label{HamJacTanV}
\end{eqnarray}
that we choose for the real vectors of the NP tetrad since they
satisfy the appropriated relations in \eqref{NPnormOrth}. In
particular, the normalisation factor for $n^\mu$ was chosen to
satisfy the normalisation condition $\ell\cdot n=1$.

We can now check that \eqref{HamJacTanV} are indeed null geodesic
generators and in our way we find Carter's constant of motion for
the NHEK. With the NP tetrad choice \eqref{NPtetrad} we find the
Weyl scalars \eqref{WeylScalarsNH} and NHEK is then Petrov type D.
In such a spacetime, if we take
$k^\mu=(\dot{t},\dot{r},\dot{\theta},\dot{\varphi})$ to be an
affinely parametrised geodesic, $k^\mu\nabla_\mu k_\nu=0$, then
\cite{ChandrasekharBook}
\begin{eqnarray}
  K &=& 2|\Psi_2|^{-2/3} ({\bf k}\cdot \bm{\ell})({\bf k}\cdot{\bf n})-Q|{\bf
k}|^2  \nonumber \\
    &=& 2|\Psi_2|^{-2/3} ({\bf k}\cdot{\bf m})({\bf k}\cdot{\bf
\overline{m}})-\lp Q-|\Psi_2|^{-2/3}\rp |{\bf k}|^2 \,,
 \label{CarterTh}
\end{eqnarray}
is conserved along $k$ if and only if a scalar $Q$ exists which
satisfies the equations
\eq
 D Q= D |\Psi_2|^{-2/3}\,, \qquad \Delta Q=\Delta |\Psi_2|^{-2/3}\,, \qquad
\delta Q=\delta^* Q=0 \,.
  \label{CarterTh2}
\eeq
In our case, from \eqref{WeylScalarsNH}, one has
$|\Psi_2|^{-2/3}=M^{4/3}\lp 1+\cos^2\theta \rp$. The scalar
$Q=M^{4/3}$ satisfies \eqref{CarterTh2} and we can then construct
the two conserved Carter quantities \eqref{CarterTh}. This yields
the pair of equations
\begin{eqnarray}
 && M^2 \lp 1+\cos^2\theta \rp^2 \dot{\theta}^2=
-L^2\,\frac{\lp 1+\cos^2\theta \rp^2}{4\sin^2\theta}-M^2\lp \delta
\cos^2\theta-K \rp , \nonumber \\
 && M^2 \lp 1+\cos^2\theta \rp^2 \dot{r}^2=  \lp
E-rL\rp^2-M^2(\delta +K)(1+r^2)\,,
 \label{CarterEqs}
\end{eqnarray}
whose RHS is, respectively, $\Theta(\theta)$ and $\mathcal{R}(r)$
defined in \eqref{HamJacEqs} if we identify $\Lambda\equiv M^2 K$.
 Carter's equations \eqref{CarterEqs}, combined with the energy and
 angular momentum conservation equations \eqref{EnAngMomCons}, reduce
 the finding of geodesics in NHEK to a quadrature problem.

If we want shear-free null geodesics we demand $\delta=0$ and
$\dot{\theta}=0$ which implies the relation \eqref{HJcondition1}.
The radial equation then stays
\eq
 M^2 \lp 1+\cos^2\theta_0 \rp^2 \dot{r}^2=
 \lp E-rL\rp^2-(1+r^2 )L^2\,\frac{\lp 1+\cos^2\theta_0
 \rp^2}{4\sin^2\theta_0}\,,
  \label{CarterRad}
\eeq
which can be independent of $r$ only for $L=0$. Inserting this
condition in \eqref{EnAngMomCons} yields \eqref{EnAngMomCons2}, and
under the redefinition $E\rightarrow E \lp 1+\cos^2\theta_0 \rp$  we
finally confirm that shear-free null geodesics are those that
satisfy
\eq
 \dot{t}=\frac{E}{1+r^2}\,,\qquad \dot{r}=\pm E\,,\qquad
 \dot{\theta}=0\,,\qquad \dot{\varphi}=-\frac{r}{1+r^2}\,E\,.
  \label{NullGeod}
\eeq
These two geodesics give us the null NP vectors $\bm{\ell}$ and
${\bf n}$ as well as the Eddington-Finkelstein coordinates for the
NHEK.  The NP tetrad is completed with the introduction of the
complex conjugate pair of vectors $m^\mu$ and $\overline{m}^\mu$ as
defined in \eqref{NPtetrad}. These are found requiring that the NP
tetrad satisfies the normalization and orthogonality conditions
\begin{eqnarray}
 && \bm{\ell} \cdot {\bf m} =\bm{\ell} \cdot {\bf \overline{m}} ={\bf n} \cdot
{\bf
m} ={\bf n} \cdot {\bf \overline{m}} =0, \nonumber\\
 && \bm{\ell} \cdot \bm{\ell}
={\bf n} \cdot {\bf n} ={\bf m} \cdot {\bf m} ={\bf \overline{m}}
\cdot {\bf \overline{m}} =0, \nonumber\\
 && \bm{\ell} \cdot {\bf n} =1,\,\,\,\,\,\,\, {\bf m} \cdot {\bf \overline{m}}
=-1. \label{NPnormOrth}
\end{eqnarray}

In terms of the NP tetrad, the metric components read
\eq g_{\mu\nu}= 2\ell_{(\mu} n_{\nu)} - 2 m_{(\mu}
\overline{m}_{\nu)}\,. \label{NPmetric}\eeq

The 12 complex spin coefficients are introduced through linear
combinations of the 24 Ricci rotation connection coefficients
$\gamma_{cab} = e_{(c)}^{\:\:\:\: \mu} e_{(b)}^{\:\:\:\: \nu}
\nabla_\nu e_{(a)\,\mu}$,
\begin{eqnarray}
\hspace{-0.5cm} && \kappa=\gamma_{311}=0, \qquad
\sigma=\gamma_{313}=0, \qquad \nu=\gamma_{242}=0, \qquad
\lambda=\gamma_{244}=0, \qquad
\epsilon=\frac{1}{2} (\gamma_{211}+\gamma_{341})=0, \nonumber\\
\hspace{-0.5cm}  &&
 \mu=\gamma_{243}=0, \qquad \rho=\gamma_{314}=0, \qquad
\gamma=\frac{1}{2} (\gamma_{212}+\gamma_{342})=\frac{r}{2M^2\lp
1+\cos^2\theta\rp},
  \nonumber\\
\hspace{-0.5cm}  &&
  \tau=\gamma_{312}=
 -\frac{i \sin\theta}{\sqrt{2} M\lp 1+\cos^2\theta\rp}, \qquad
 \alpha=\frac{1}{2} (\gamma_{214}+\gamma_{344})=
 -\frac{\cos\theta -i\lp 2-\cos^2\theta\rp }{2\sqrt{2} M\lp
1-i\,\cos\theta\rp^2\sin\theta}, \nonumber\\
\hspace{-0.5cm}  &&
 \pi=\gamma_{241}=
 \frac{i \sin\theta}{\sqrt{2} M\lp 1-i\,\cos\theta\rp^2}, \qquad
\beta=\frac{1}{2} (\gamma_{213}+\gamma_{343})=
 \frac{\cos\theta}{2\sqrt{2} M\lp 1+i\,\cos\theta\rp
\sin\theta}\,.\label{spincoef}
\end{eqnarray}
Their complex conjugates are obtained through the replacement
$3\leftrightarrow 4$ in $\gamma_{cab}$. From the Goldberg-Sachs
theorem, $\kappa=\sigma=\nu=\lambda=0$ implies that the NHEK is
Petrov type D (as it must be by construction). Moreover,
$\epsilon=0$ implies that $\bm{\ell}$ is affinely parametrised as it
is indeed the case.

The Weyl tensor
\eq C_{\mu\nu\alpha\beta}=R_{\mu\nu\alpha\beta} -\frac{1}{2}\lp
g_{\mu\alpha}R_{\nu\beta} + g_{\nu\beta} R_{\mu\alpha}
-g_{\nu\alpha}R_{\mu\beta} - g_{\mu\beta} R_{\nu\alpha}\rp +
\frac{1}{6}\lp g_{\mu\alpha}g_{\nu\beta}-g_{\mu\beta}g_{\nu\alpha}
\rp ,\eeq
reduces to the Riemann tensor because \eqref{NHkerr} is Ricci flat.
 The 5 complex Weyl scalars $\Psi_i$ in the NP formalism encode the
information on the 10 independent components $C_{abcd}$ of the Weyl
tensor,
%
\eqn && \Psi_0 = -C_{1313} =  - C_{\mu\nu\alpha\beta}\, \ell^\mu
m^\nu
\ell^\alpha m^\beta, \nonumber \\
 && \Psi_1 = -C_{1213} = - C_{\mu\nu\alpha\beta}\, \ell^\mu n^\nu \ell^\alpha
m^\beta, \nonumber \\
 && \Psi_2 = -C_{1342} = - C_{\mu\nu\alpha\beta}\, \ell^\mu  m^\nu
\overline{m}^\alpha n^\beta, \nonumber \\
 && \Psi_3 = -C_{1242} = - C_{\mu\nu\alpha\beta}\, \ell^\mu n^\nu
\overline{m}^\alpha n^\beta, \nonumber \\
 && \Psi_4 = -C_{2424} = - C_{\mu\nu\alpha\beta}\, n^\mu \overline{m}^\nu
n^\alpha \overline{m}^\beta\,.\label{WeylScalars}
 \eeqn
For the NHEK these Weyl scalars are listed in \eqref{WeylScalarsNH}.

The fundamental quantities in the NP formalism needed to study
perturbations are the spin coefficients listed in \eqref{spincoef}
and the directional derivative operators,
\eq D=\ell^\mu\nabla_\mu\,,\qquad \Delta=n^\mu\nabla_\mu\,,\qquad
\delta=m^\mu\nabla_\mu\,,\qquad
\delta^*=\overline{m}^\mu\nabla_\mu\,. \label{DirectDeriv}\eeq
%

\subsection{Master equation for NHEK in
Poincar\'e coordinates \label{sec:MasterPoincare}}

For completeness we write here the master equation for NHEK in
Poincar\'e coordinates. This is the counterpart of the global
coordinate master equation \eqref{MasterEq}.

Let quantities with tildes denote Boyer-Lindquist coordinates of the
full black hole solution \eqref{kerr}, and take $\{ \tau, y, \theta,
\varphi \}$ to be the Poincar\'e coordinates describing NHEK.

Bardeen and Horowitz define the near-horizon limit of the extreme
Kerr solution by setting \cite{Bardeen:1999px} \eq
 \tilde{r} = a + \lambda y, \qquad \tilde{t} = \frac\tau{\lambda}, \qquad \tilde{\phi} = \varphi + \frac\tau{2a\lambda},
\eeq
 where $a$ is the extreme value for the Kerr rotation parameter, and taking the limit
$\lambda \rightarrow 0$ with the untilded quantities held fixed. The
limit yields the near-horizon solution in Poincar\'e coordinates.
Taking this limit in the Kinnersley tetrad \cite{Teukolsky:1973ha},
one finds that
 \begin{eqnarray} && \lambda \ell \rightarrow
\frac{2a^2}{y} \frac{\partial}{\partial \tau} +
\frac{\partial}{\partial y} - \frac{1}{y} \frac{\partial}{\partial
\varphi}\,, \nonumber \\
 && \lambda^{-1} n \rightarrow \frac{1}{a^2 (1+\cos \theta)} \left( a^2 \frac{\partial}{\partial \tau}
 - \frac{y^2}{2} \frac{\partial}{\partial y} - \frac{y}{2} \frac{\partial}{\partial \varphi}
\right),
 \nonumber \\
 &&
 m \rightarrow \frac{1}{\sqrt{2} a (1+i \cos \theta)} \left( \frac{\partial}{\partial \theta}
+ i \frac{(1+\cos^2\theta)}{2 \sin \theta} \frac{\partial}{\partial
\varphi} \right).
 \end{eqnarray}
 Hence, by performing a boost before
taking the limit, we can ensure that the tetrad remains well-defined
and must therefore give a tetrad aligned with the principal null
directions of the near-horizon geometry.

Consider the Teukolsky equation for a field $\psi$ of spin $s$ in
the extreme Kerr geometry \cite{Teukolsky:1973ha}. Let \eq
 \psi = f(\tau,y,\theta,\varphi) = f\lp\lambda \tilde{t}, \frac{\tilde{r}-a}{\lambda}, \theta,\tilde{\phi} - \frac{\tilde{t}}{2a} \rp,
\eeq Plugging this into the Teukolsky master equation
\cite{Teukolsky:1973ha}, and taking $\lambda \rightarrow 0$, we find
that it becomes
 \eqn \label{nhteuk1} &&
\frac{4a^4}{y^2} \partial_\tau^2 f - \frac{4a^2}{y}
\partial_\tau
\partial_\varphi f + \left(2 - \frac{1}{4} \sin^2 \theta -
\frac{1}{\sin^2 \theta} \right) \partial_\varphi^2 f - y^{-2s}
\partial_y \left( y^{2s+2} \partial_y f \right) - \frac{1}{\sin \theta} \partial_\theta
\left( \sin \theta \partial_\theta f\right) \nonumber \\
& & \hspace{1cm} -is \left( \frac{2\cos\theta}{\sin^2 \theta} +
\cos\theta \right)
\partial_\varphi f - \frac{4a^2 s}{y} \partial_\tau f + (s^2 \cot^2 \theta
-s ) f=0 \eeqn This is the master equation governing perturbations
of the near-horizon Kerr geometry written in Poincar\'e coordinates.

We separate variables by setting \eq
 f(\tau,y,\theta,\varphi) = F(\tau,y) S(\theta) e^{im\varphi}.
\eeq Equation (\ref{nhteuk1}) separates into an angular equation
given by \eqref{AngEq} with $C=m/2$ and into \eq \label{treq1}
 \frac{4a^4}{y^2} \partial_\tau^2 F - \frac{4a^2(s+im)}{y}
  \partial_\tau  F - y^{-2s} \partial_y \left( y^{2s+2}
\partial_y F \right) + \left(\Lambda^{(s)}_{lm}-\frac{7m^2}{4}\right) F=0,
\eeq
 where $\Lambda^{(s)}_{lm}$ is the constant of separation already
discussed after \eqref{AngEq}.

\setcounter{equation}{0}\section{Phase and group velocities
\label{sec:veloc}}

In this Appendix we give some details of the analysis done in
subsection \ref{sec:travel}.

To discuss travelling waves and their phase and group velocities we
need the next-to-leading contribution to the asymptotic behavior
\eqref{RadialSolAsymp}. This is obtained applying to
\eqref{RadialSol} the transformation law $z\rightarrow 1/(1-z)$ of
the hypergeometric function and its asymptotic expansion for large
radial distances yielding,
\begin{eqnarray}
  \Phi^{(s)}_{lm\omega} &\approx & 2^{(1+\eta)/2}\Gamma(b-a) C^\pm e^{\pm i \pi(\beta
  -\alpha -a)/2} e^{-\frac{1+\eta}{2}\ln|r|-\frac{2 q \omega}{1+\eta}\frac{1}{r} }
   \nonumber\\
   && + 2^{(1-\eta)/2}\Gamma(a-b)
  D^{\pm} e^{\pm i \pi(\beta-\alpha -b)/2} e^{-\frac{1-\eta}{2}\ln|r|-\frac{2 q \omega}{1-\eta}\frac{1}{r} },
\label{RadialSolAsympNext}
 \end{eqnarray}
where the amplitudes  $C^\pm$ and $D^\pm$ are defined in
\eqref{amplitudes}.

Introduce the quantities
\begin{eqnarray}
 &&\hspace{-0.5cm} S_\pm(r)=\exp{ \left[\, i\lp \pm \frac{1}{2}\,\widetilde{\eta} \ln|r|
 +\frac{2\omega\lp \mp m\,\widetilde{\eta}+s
\rp}{1+\widetilde{\eta}^{\,2}}\frac{1}{r}   \rp \right]} \,, \qquad
\eta=i\widetilde{\eta}\,, \nonumber\\
 &&\hspace{-0.5cm} k_\pm(r)=-i\,\frac{\partial_r S_\pm(r)}{S_\pm(r)}
 \,.
  \label{AuxVeloc}
\end{eqnarray}
Here, $S_\pm(r)$ encodes the radial contribution to the wave
propagation and, in a WKB approximation to the traveling waves,
$k_\pm(r)$ is the effective wavenumber of the wave. The superscript
'$+$' ('$-$') is in correspondence with the amplitude $C^\pm$
($D^\pm$). The phase velocity of the traveling wave is then
 \eq
 v_{\rm ph}^{\pm}= \frac{\omega}{k_\pm(r)} \simeq \pm \frac{2\omega \,r}{\widetilde{\eta}}\,,
  \label{PhVeloc}
 \eeq
while the group velocity is
 \eq
 v_{\rm g}^{\pm }= \lp\frac{d k_\pm(r)}{d\omega}\rp^{-1}
 =\pm  \frac{1}{2}\, \frac{(1+\widetilde{\eta}^2)\, r^2}{m\,\widetilde{\eta} \mp s } \,.
  \label{GrVeloc}
 \eeq
For $2\leq l \leq 20 $ and the values of $|m|\leq l$ that yield
$\widetilde{\eta}>0$  we have checked that, in the denominator of
$v_{\rm g}^{\pm }$, $m\,\widetilde{\eta} \mp s$ is positive for
$m>0$ and negative for $m<0$.

Both at $r=\pm\infty$, depending on whether we choose the $C^\pm$ or
the $D^\pm$ contributions in \eqref{RadialSolAsympNext}, we can have
the combinations for the sign of the phase and group velocities
displayed in Table \ref{Table:Veloc}.
\begin{table}[bh]
\begin{eqnarray}
\nonumber
\begin{array}{|||c|c|||c|c|c|c|||}\hline\hline\hline
 \,  &\,  &  C^-  & D^- &
   C^+ & D^+ \\
\hline \hline\hline
 v_{\rm ph}& {\rm Re}(\omega)>0 & >0 & <0 & <0 & >0 \\
 \hline
 v_{\rm ph}& {\rm Re}(\omega)<0 & <0 & >0 & >0 & <0 \\
 \hline\hline
 v_{\rm g}& m>0 & <0 & >0 & <0 & >0 \\
 \hline
 v_{\rm g}& m<0 & >0 & <0 & >0 & <0 \\
\hline\hline\hline
\end{array}
\end{eqnarray}
\caption{Phase and group velocities for the possible amplitudes
choices in the asymptotic solution \eqref{RadialSolAsymp}.}
\label{Table:Veloc}
\end{table}
%

\setcounter{equation}{0}
 \section{Hertz map between Weyl scalar and metric perturbations}
 \label{sec:Wald}

In this Appendix we derive the map $h_{\mu\nu}(\Psi)$ between the
Weyl scalars $\Psi_{0,4}$ (that satisfy the decoupled Teukolsky
equations) and the metric perturbations $h_{\mu\nu}$, as well as the
map $A_\mu(\phi)$ between the NP complex scalars perturbations
$\phi_{0,2}$ and the Maxwell perturbed vector potential $A_\mu$. We
follow \cite{Wald} and recover the results of
\cite{Cohen:1975,Chrzanowski:1975,Cohen:1979,Wald,Stewart1979} but
we take the opportunity to emphasize the importance of these works
(not so well-known in the community) and to give some details not
presented in the original articles and to also pinpoint small typos
in some of these works that have propagated in the literature. We
start by reviewing Wald's work \cite{Wald} in the next subsection.
Then we apply it to get $h_{\mu\nu}(\Psi)$ (subsection
\ref{sec:WaldGravity}), and to find $A_\mu(\phi)$ (subsection
\ref{sec:WaldElectromag}). A similar analysis could be done to
obtain the map for the Weyl fermionic perturbations.

 \subsection{Problem statement. The Hertz potential map}
 \label{sec:WaldProof}

Let us start by briefly reviewing the seminal work of Teukolsky
\cite{Teukolsky:1972my,Teukolsky:1973ha}. Suppose we wish to solve
the perturbation equation $\varepsilon(h)=0$ where $\varepsilon$ is
a linear differential operator and $h$ is the field perturbation on
which $\varepsilon$ acts. For example, $\varepsilon$ can be the
Maxwell operator describing electromagnetic perturbations
$A_\alpha$, $\left[\Delta_{\rm E}(A_\alpha)\right]_\mu=0$, \ie
$\nabla_\nu F^{\mu\nu}=0$; or it can be the gravitational operator
describing gravitational perturbations $h_{\alpha\beta}$,
$\left[\Delta_{\rm G}(h_{\alpha\beta})\right]_{\mu\nu}=0$ written in
\eqref{Lichnerowicz}.

Suppose now that:
\begin{itemize}
\item a new variable $\Psi=L_\Psi(h)$, where $L_\Psi$ is a linear
differential operator, has been introduced (these are \eg the NP
complex electromagnetic scalars $\phi_{0,1,2}$ or the Weyl scalars
$\Psi_{0,\cdots,4}$);
\item a linear partial differential operator $\mathbb{D}$ has been
found such that for all $h$ one has
\eq
 \mathbb{D}\varepsilon(h)=\mathcal{O}L_\Psi(h)=\mathcal{O}\Psi\,,
  \label{TeukHuvToWeylAux}
\eeq
where $\mathcal{O}$ is another partial differential operator.
\end{itemize}
Then
\eq
 \varepsilon(h)=0 \quad \Rightarrow \quad \mathcal{O}(\Psi)=0\,.
  \label{TeukHuvToWeyl}
\eeq
This is the main result of \cite{Teukolsky:1972my,Teukolsky:1973ha},
who found the variable $\Psi$ and operators $L_\Psi$, $\mathbb{D}$
and $\mathcal{O}$, as well as the well-known decoupled Teukolsky
equations \eqref{TeukHuvToWeylAux}, that describe the problem of
electromagnetic, Weyl fermionic and gravitational perturbations in
the Kerr black hole.

So assume that we have carried on the previous steps that lead to
\eqref{TeukHuvToWeyl} and that furthermore we solved it and have a
solution for $\Psi$. The next question is how to get the original
perturbation $h$ from the knowledge of the scalar perturbation
$\Psi$, \ie the unique map $h=h(\Psi)$. This issue has been
addressed by Cohen and Kegeles \cite{Cohen:1975,Cohen:1979}, and
Chrzanowski \cite{Chrzanowski:1975} and later Wald \cite{Wald}
proved rigorously and in a few lines their results. In the sequel we
review Wald's proof \cite{Wald}.

Start by recalling the notion of adjoint of an operator. Let
$\mathcal{O}$ be a linear differential operator taking a scalar,
vectorial or tensorial field into another similar field. Then there
is an unique adjoint operator $\mathcal{O}^\dag$ such that
\eq
 \Psi_{\rm H} \lp \mathcal{O} \Psi\rp - \lp \mathcal{O}^\dag \Psi_{\rm H} \rp
\Psi=\nabla_\mu s^\mu
 \,,
  \label{adjoint}
\eeq
for arbitrary fields $\Psi$ and $\Psi_{\rm H}$, and where
$\nabla_\mu s^\mu$ is a total divergence.

Wald's theorem states the following. Assume that:
\begin{itemize}
\item the identity \eqref{TeukHuvToWeylAux} is satisfied for the
linear differential operators $\mathbb{D},\,\varepsilon,\,
\mathcal{O},\,L_\Psi$;
\item $\Psi_{\rm H}$ satisfies $\mathcal{O}^\dag \Psi_{\rm H}=0$, where
$\Psi_{\rm H}$ is called the Hertz potential;
\end{itemize}
Then
\begin{itemize}
\item $\mathbb{D}^\dag \Psi_{\rm H}$ satisfies $\varepsilon^\dag \lp
\mathbb{D}^\dag \Psi_{\rm H} \rp=0$.
\item Moreover, this in particular also implies that if
$\varepsilon$ is self-adoint, $\varepsilon^\dag = \varepsilon$, then
\eq
 h=\mathbb{D}^\dag \Psi_{\rm H} \:\: {\rm is} \:\: {\rm a} \:\: {\rm solution}
\:\: {\rm
 of} \:\: \varepsilon(h)=0\,.
  \label{WaldMap}
\eeq
\end{itemize}
and provides the map we are looking for.

The proof of this result is short and simple \cite{Wald}. Taking the
adjoint of $\mathbb{D}\varepsilon=\mathcal{O}L_\Psi$ one has
$\varepsilon^\dag \mathbb{D}^\dag = L_\Psi^\dag \mathcal{O}^\dag$.
Applying these operators to $\Psi_{\rm H}$ one gets
$\varepsilon^\dag \mathbb{D}^\dag \Psi_{\rm H}=0$ after using the
assumption $\mathcal{O}^\dag \Psi_{\rm H}=0$. Moreover, if
$\varepsilon^\dag = \varepsilon$, it trivially follows that
$\varepsilon \lp \mathbb{D}^\dag \Psi_{\rm H}\rp=0$, \ie
$h=\mathbb{D}^\dag \Psi_{\rm H}$ is a solution of the initial
perturbation equation $\varepsilon(h)=0$.

Finally note that the Hertz potential construction reviewed here
applies to geometries where no energy-momentum tensor is present.
For a discussion of the method when this is not the case as well as
for the second order perturbation analysis we ask the reader to see
\cite{Price:2006ke}.

 \subsection{Application: the map $ h_{\mu\nu}(\Psi)$ for gravitational
perturbations}
 \label{sec:WaldGravity}

Our starting point are the decoupled Teukolsky equations for the
perturbed Weyl scalars $\Psi_0^{(1)}$ and $\Psi_4^{(1)}$, namely
equations (2.12)-(2.15) of \cite{Teukolsky:1973ha}, which are
written in \eqref{TeukPositSpin} for $s=2$ and \eqref{TeukNegSpin}
for $s=-2$. These can be written as (to make the connection with the
nomenclature of the previous subsection straightforward)
\begin{eqnarray}
&& \mathcal{O}_{\rm G_0}\Psi_0^{(1)} =  \mathbb{D}_{\rm
G_0}^{\mu\nu}
T_{\mu\nu}^{(1)}\,, \nonumber\\
 && \mathcal{O}_{\rm G_4}\Psi_4^{(1)} = \mathbb{D}_{\rm G_4}^{\mu\nu}
T_{\mu\nu}^{(1)}\,,
  \label{TeukDecouple}
\end{eqnarray}
where we use the superscript $(1)$ to denote a perturbed quantity
(otherwise it refers to an unperturbed quantity), and
\begin{eqnarray}
\hspace{-0.9 cm}\mathcal{O}_{\rm G_0}\!\!\!&=& \!\!\!
(D-3\epsilon+\overline{\epsilon}-4\rho-\overline{\rho})
 (\Delta+\mu-4 \gamma)-(\delta+\overline{\pi}-\overline{\alpha} -3\beta-4\tau)
(\overline{\delta}+\pi-4\alpha)-3\Psi_{2} \,, \nonumber\\
 \hspace{-0.9 cm} \mathcal{O}_{\rm G_4}\!\!\!&=& \!\!\! (\Delta
+3\gamma-\overline{\gamma}+4\mu+\overline{\mu})(D+4\epsilon-\rho)-(\overline{\delta}
 -\overline{\tau}+\overline{\beta}+3\alpha+4\pi)(\delta-\tau+4\beta)-3\Psi_{2}
 \,,
  \label{TeukOperatorO}
\end{eqnarray}
and
\begin{eqnarray}
\hspace{-1 cm} \mathbb{D}_{\rm G_0}^{\mu\nu}
&=&(\delta+\overline{\pi}-\overline{\alpha}-3\beta-4\tau)\left[(D-2\epsilon-2\overline{\rho})\ell^{(\mu}m^{\nu)}
-(\delta+\overline{\pi}-2\overline{\alpha}-2\beta)\ell^\mu\ell^\nu\right]
\nonumber \\
&&+(D-3\epsilon+\overline{\epsilon}-4\rho-\overline{\rho})
\left[(\delta+2\overline{\pi}-2\beta)\ell^{(\mu}m^{\nu)}-(D-2\epsilon+2\overline{\epsilon}-\overline{\rho})m^\mu
m^\nu\right],\nonumber \\
 \hspace{-1 cm} \mathbb{D}_{\rm G_4}^{\mu\nu}
  &=&
 (\Delta +3\gamma-\overline{\gamma}+4\mu+\overline{\mu})
 \left[(\overline{\delta} -2\overline{\tau}+2\alpha)
n^{(\mu}\overline{m}^{\nu)}
 -(\Delta +2\gamma-2\overline{\gamma}+\overline{\mu})
 \overline{m}^\mu\overline{m}^\nu\right] \nonumber \\
&&
+(\overline{\delta}-\overline{\tau}+\overline{\beta}+3\alpha+4\pi)
\left[ (\Delta+2\gamma+2\overline{\mu})n^{(\mu}\overline{m}^{\nu)}
 -(\overline{\delta}-\overline{\tau}+2\overline{\beta}+2\alpha)n^\mu n^\nu
 \right]\,. \label{TeukOperatorD}
\end{eqnarray}

Following Wald's procedure, the Hertz potential $\Psi_{\rm H}$ is
introduced to be such that it satisfies the equation
$\mathcal{O}_{\rm G}^\dag\Psi_{\rm H}=0$, that is
\begin{eqnarray}
\hspace{-1 cm} &&
\left[(\Delta+3\gamma-\overline{\gamma}+\overline{\mu})(D+4\epsilon+3\rho)
 -
(\overline{\delta}+\overline{\beta}+3\alpha-\overline{\tau})(\delta+4\beta+3\tau)-3\Psi_{2}\right]\Psi_{\rm
H_0}=0 \,, \nonumber\\
\hspace{-1 cm} && \left[
 (D-3\epsilon+\overline{\epsilon}-\overline{\rho}) (\Delta -4\gamma-3\mu)
  -(\delta-3\beta-\overline{\alpha}+\overline{\pi})
(\overline{\delta}-4\alpha-3\pi) -3\Psi_{2}\right]\Psi_{\rm
  H_4}=0 \,,
  \label{TeukOperatorOdag}
\end{eqnarray}
where, to obtain the adjoint of \eqref{TeukOperatorO}, we used the
relations\footnote{To get \eqref{BasicAdjoints}, introduce the
internal product $\langle \Psi_{\rm H},\mathcal{O}\psi\rangle
=\int_a^b d^4x\sqrt{-g}\,\Psi_{\rm H}\mathcal{O}\psi$. For the
directional derivative operators
$\mathcal{O}=e_{(a)}^{\:\:\:\:\mu}\nabla_\mu$ it then follows, after
integrations by parts and use of $\nabla_\mu \sqrt{-g}=0$, that
$\langle \Psi_{\rm H},\mathcal{O}\psi\rangle =-\langle \psi
,\mathcal{O}^\dag \Psi_{\rm H} \rangle + \int_a^b \nabla_\mu s^\mu$
with $\mathcal{O}^\dag = -\mathcal{O} -\nabla_\mu
 e_{(a)}^{\:\:\:\:\mu}$,
and $\int_a^b \nabla_\mu s^\mu$ being the short notation for a total
divergence contribution. To compute $\nabla_\mu
 e_{(a)}^{\:\:\:\:\mu}$ use the well-known relation between
 covariant derivative of the NP tetrad and the spin coefficients
\cite{ChandrasekharBook},
 $\nabla_\mu e_{(a)\,\nu} = e^{(c)}_{\:\:\:\:\nu} \gamma_{cab}
 e^{(b)}_{\:\:\:\:\mu}$, and relations \eqref{spincoef}.
 }
\begin{eqnarray}
&& D^\dag=-(D+\epsilon+\overline{\epsilon}-\rho-\overline{\rho}) \,,
\qquad
\Delta^\dag=-(\Delta-\gamma-\overline{\gamma}+\mu+\overline{\mu})
\,, \nonumber\\
&& \delta^\dag=-(\delta+\beta-\overline{\alpha}-\tau+\overline{\pi})
\,, \qquad
\overline{\delta}^\dag=-(\overline{\delta}+\overline{\beta}-\alpha-\overline{\tau}+\pi)
\,,
  \label{BasicAdjoints}
\end{eqnarray}
and the well known property $(AB)^\dag =B^\dag A^\dag$. Equations
 \eqref{TeukOperatorOdag} can be written, respectively,  as
\eqref{HertzPotential:NegSpin} with $s=-2$ and
\eqref{HertzPotential:PosSpin} with $s=2$.

Since the gravitational perturbation operator
$\varepsilon\equiv\left[\Delta_{\rm
G}(h_{\alpha\beta})\right]_{\mu\nu}$ is self-adjoint, Wald's theorem
tell us that the map \eqref{WaldMap} between the Hertz potential
$\Psi_{\rm H}$ and the metric perturbations $h_{\mu\nu}$ is given by
$h_{\mu\nu}=2{\rm Re}\left[ \mathbb{D}^\dag_{\rm G
\:\mu\nu}\Psi_{\rm H}\right]$, \ie
\begin{eqnarray}
h_{\mu\nu}^{IRG}\!\!\!\!&=& \!\!\!\!{\biggl \{} \ell_{(\mu}m_{\nu)}
\left[ (D+3\epsilon+\overline{\epsilon}-\rho+\overline{\rho})
(\delta+4\beta+3\tau)
 +(\delta+3\beta-\overline{\alpha}-\tau-\overline{\pi})(D+4\epsilon+3\rho)
 \right]  \nonumber \\
 && -\ell_\mu \ell_\nu (\delta+3\beta+\overline{\alpha}-\overline{\tau})
(\delta+4\beta+3\tau) -m_\mu m_\nu
(D+3\epsilon-\overline{\epsilon}-\rho) (D+4\epsilon+3\rho) {\biggl
\}}\Psi_{\rm
H_0}+{\rm c.c.}\,, \nonumber\\
h_{\mu\nu}^{ORG}\!\!\!\!&=&\!\!\!\! {\biggl \{}
n_{(\mu}\overline{m}_{\nu)} \left[
(\overline{\delta}+\overline{\beta}-3\alpha+\overline{\tau}+\pi)
(\Delta-4\gamma -3\mu)
 +(\Delta-3\gamma-\overline{\gamma}+\mu-\overline{\mu})(\overline{\delta}-4\alpha-3\pi)
 \right] \nonumber \\
 && \hspace{-0.5cm}-n_\mu n_\nu (\overline{\delta}-\overline{\beta}-3\alpha+\pi)
(\overline{\delta}-4\alpha -3\pi) -\overline{m}_\mu \overline{m}_\nu
(\Delta-3\gamma+\overline{\gamma}+\mu) (\Delta-4\gamma-3\mu) {\biggl
\}}\Psi_{\rm H_4}+{\rm c.c.}\,, \nonumber \\
&&
  \label{MapHertzHuv}
\end{eqnarray}
where to get the adjoint of \eqref{TeukOperatorD} we used again
\eqref{BasicAdjoints}, and ${\rm c.c.}$ stands for complex
conjugate. The first of these relations gives the metric
perturbations in the ingoing radiation gauge ($IRG$), while the
second provides the map in the outgoing radiation gauge ($ORG$); see
\eqref{gauges}. The first relation agrees with the results of
\cite{Cohen:1975,Chrzanowski:1975,Cohen:1979,Wald}, and is equation
\eqref{huvHertz} in the main body of the text. The outgoing
radiation map corrects typos in the relation of Table 1 of
\cite{Chrzanowski:1975} that have propagated in the literature.

 \subsection{Application: the map $ A_\mu(\phi)$ for electromagnetic
perturbations}
 \label{sec:WaldElectromag}

We begin our discussion with the decoupled Teukolsky equations for
the perturbed electromagnetic NP scalars $\phi_0^{(1)}$ and
$\phi_2^{(1)}$, namely equations (3.5)-(3.8) of
\cite{Teukolsky:1973ha}, which are written in \eqref{TeukPositSpin}
for $s=1$ and \eqref{TeukNegSpin} for $s=-1$. These can be written
as
\begin{eqnarray}
&& \mathcal{O}_{\rm E_0}\phi_0^{(1)} =  \mathbb{D}_{\rm
E_0}^\mu J_\mu^{(1)}\,, \nonumber\\
 && \mathcal{O}_{\rm E_2}\phi_2^{(1)} = \mathbb{D}_{\rm E_2}^\mu J_\mu^{(1)}\,,
  \label{TeukDecouple:Max}
\end{eqnarray}
where, again, we use the superscript $(1)$ to denote a perturbed
quantity, and
\begin{eqnarray}
\hspace{-1 cm} && \mathcal{O}_{\rm E_0}=
(D-\epsilon+\overline{\epsilon}-2\rho-\overline{\rho})
 (\Delta+\mu-2\gamma)-(\delta-\beta-\overline{\alpha}-2\tau+\overline{\pi})
(\overline{\delta}+\pi-2\alpha)\,, \nonumber\\
\hspace{-1 cm} && \mathcal{O}_{\rm E_2}= (\Delta
+\gamma-\overline{\gamma}+2\mu+\overline{\mu})(D+2\epsilon-\rho)-(\overline{\delta}+\alpha+\overline{\beta}
 +2\pi-\overline{\tau})(\delta-\tau+2\beta)
 \,,
  \label{TeukOperatorO:Max}
\end{eqnarray}
and
\begin{eqnarray}
&&\mathbb{D}_{\rm E_0}^\mu
=(\delta-\overline{\alpha}-\beta+\overline{\pi}-2\tau)\ell^\mu
-(D-\epsilon+\overline{\epsilon}-2\rho-\overline{\rho})m^\mu
\,,\nonumber\\
&& \mathbb{D}_{\rm E_2}^\mu =
 (\Delta +\gamma-\overline{\gamma}+2\mu+\overline{\mu})\overline{m}^{\mu}
 -(\overline{\delta}+\alpha+\overline{\beta}+2\pi-\overline{\tau})n^\mu \,.
\label{TeukOperatorD:Max}
\end{eqnarray}

The Hertz potential $\Psi_{\rm H}$ is introduced to be such that it
satisfies the equation $\mathcal{O}_{\rm E}^\dag\Psi_{\rm H}=0$,
that is
\begin{eqnarray}
&&
\left[(\Delta+\gamma-\overline{\gamma}+\overline{\mu})(D+2\epsilon+\rho)
 -
(\overline{\delta}+\overline{\beta}+\alpha-\overline{\tau})(\delta+2\beta+\tau)
\right]\Psi_{\rm H_0}=0 \,, \nonumber\\
&& \left[
 (D-\epsilon+\overline{\epsilon}-\overline{\rho}) (\Delta -2\gamma-\mu)
  -(\delta-\beta-\overline{\alpha}+\overline{\pi})
(\overline{\delta}-2\alpha-\pi)\right]\Psi_{\rm
  H_2}=0 \,,
  \label{TeukOperatorOdag:Max}
\end{eqnarray}
where, to obtain the adjoint of \eqref{TeukOperatorO:Max}, we used
\eqref{BasicAdjoints} and $(AB)^\dag =B^\dag A^\dag$. Equations
\ref{TeukOperatorOdag:Max} are, respectively, equations
\eqref{HertzPotential:NegSpin} for $s=-1$ and
\eqref{HertzPotential:PosSpin} for $s=1$.

Since the Maxwell perturbation operator
$\varepsilon\equiv\left[\Delta_{\rm E}(A_\alpha)\right]_\mu$ is
self-adjoint, Wald's theorem tell us that the map \eqref{WaldMap}
between the Hertz potential $\Psi_{\rm H}$ and the vector potential
perturbations $A_\mu$ is given by $A_\mu=2{\rm Re}\left[
\mathbb{D}^\dag_{\rm E \:\mu}\Psi_{\rm H}\right]$, \ie
\begin{eqnarray}
&& A_\mu^{IRG} = \left[ -\ell_\mu (\delta+2\beta+\tau)
+m_\mu (D+2\epsilon+\rho)\right]\Psi_{\rm H_0}+{\rm c.c.}\,, \nonumber\\
&& A_\mu^{ORG} = \left[ - \overline{m}_\mu  (\Delta-2\gamma -\mu) +
n_\mu (\overline{\delta}-2\alpha -\pi) \right]\Psi_{\rm H_2}+{\rm
c.c.}\,,
  \label{MapHertzAu:Max}
\end{eqnarray}
where to get the adjoint of \eqref{TeukOperatorD:Max} we used again
\eqref{BasicAdjoints}. The first of these relations gives the
Maxwell perturbations in the ingoing radiation gauge, $\ell^\mu
A_\mu=0$, while the second provides the map in the outgoing
radiation gauge, $n^\mu A_\mu=0$. The first relation agrees with the
results of \cite{Cohen:1975,Chrzanowski:1975,Cohen:1979,Wald}. The
outgoing radiation map agrees with \cite{Cohen:1979} and corrects
typos in the relation of Table 1 of \cite{Chrzanowski:1975}.

 \subsection{Hertz potential for spin $s$ in a Ricci flat Petrov type D geometry}
 \label{sec:UnifyMaster}

For an arbitrary Petrov type D solution we can write the decoupled
Teukolsky equations in a single equation that depends on the spin of
the field \cite{Teukolsky:1973ha}, and the same happens for the
decoupled equations for the Hertz potential if the the solution is
furthermore Ricci flat.

For positive spin and negative spin, Teukolsky's decoupled equations
are, respectively, \cite{Teukolsky:1973ha}
\begin{eqnarray}
&& \hspace{-1cm} {\biggl [} \left[
D-(2s-1)\epsilon+\overline{\epsilon}-2s\rho-\overline{\rho}\right]
 (\Delta+\mu-2s \gamma)-\left[ \delta+\overline{\pi}-\overline{\alpha}
 -(2s-1)\beta-2s\tau \right]
(\overline{\delta}+\pi-2s \alpha)\nonumber \\
&& \hspace{2cm} -3s\lp s-\frac{1}{2}\rp (s-1)\Psi_{2}{\biggr ]}
\Psi^{(s)}=4\pi T_{(s)}, \qquad {\rm for} \quad s=+2, +1,
+\frac{1}{2}, 0\,, \label{TeukPositSpin} \\
&& \hspace{-1cm} {\biggl [} \left[ \Delta
-(2s+1)\gamma-\overline{\gamma}-2s\mu+\overline{\mu}\right]
(D-2s\epsilon-\rho)-\left[\overline{\delta}
 -\overline{\tau}+\overline{\beta}-(2s+1)\alpha-2s\pi\right] (\delta-\tau-2s \beta)\nonumber \\
&& \hspace{2cm} +3s\lp s+\frac{1}{2}\rp (s+1)\Psi_{2}{\biggr ]}
\Psi^{(s)}=4\pi T_{(s)}, \qquad {\rm for} \quad s=-2, -1,
-\frac{1}{2}, 0\,.
  \label{TeukNegSpin}
\end{eqnarray}
The relation between the nomenclature used here and the original
notation of Teukolsky \cite{Teukolsky:1973ha} was already displayed
in Table \ref{Table:Teuk} and associated discussion.

The Hertz potential $\Psi_{\rm H}^{(s)}$ in the ingoing radiation
gauge describes negative spin $s=-2,-1,-\frac{1}{2}$ field
perturbations, while in the outgoing radiation gauge $\Psi_{\rm
H}^{(s)}$ describes positive spin $s=+2,+1,+\frac{1}{2}$
perturbations. They are, respectively, the solutions of the scalar
equations
 \begin{eqnarray}
&& \hspace{-1cm} {\biggl [} \left[
\Delta-(2s+1)\gamma-\overline{\gamma}+\overline{\mu}\right]
 \left[ D-2s\epsilon-(2s+1)\rho \right]
-\left[
\overline{\delta}+\overline{\beta}-(2s+1)\alpha-\overline{\tau}\right]
\left[\delta-2s\beta-(2s+1)\tau\right]
\nonumber \\
&& \hspace{2cm} +3s\lp s+\frac{1}{2}\rp (s+1)\Psi_{2}{\biggr ]}
\Psi_{\rm H}^{(s)}=0, \qquad {\rm for} \quad s=-2, -1,
-\frac{1}{2}, 0 \,, \label{HertzPotential:NegSpin}\\
 && \hspace{-1cm} {\biggl [} \left[
D-(2s-1)\epsilon+\overline{\epsilon}-\overline{\rho}\right] \left[
\Delta -2s\gamma-(2s-1)\mu\right]
  -\left[\delta-(2s-1)\beta-\overline{\alpha}+\overline{\pi}\right]
\left[\overline{\delta}-2s\alpha-(2s-1)\pi\right] \nonumber \\
&& \hspace{2cm} -3s\lp s-\frac{1}{2}\rp (s-1)\Psi_{2}{\biggr ]}
\Psi_{\rm H}^{(s)}=0, \qquad {\rm for} \quad s=+2, +1, +\frac{1}{2},
0\,. \label{HertzPotential:PosSpin}
\end{eqnarray}
The conjugate Teukolsky perturbations $\Psi^{(s)}$ to these Hertz
potentials $\Psi_{\rm H}^{(s)}$ can be read from Tables
\ref{Table:Teuk} and \ref{Table:Teuk2}.

For the NHEK geometry written in global coordinates the NP tetrad is
written in \eqref{NPtetrad}, the spin coefficients are listed in
\eqref{spincoef}, and the directional derivative operators can be
read from \eqref{DirectDeriv}. Using this information in
\eqref{TeukPositSpin}-\eqref{TeukNegSpin}, and
\eqref{HertzPotential:NegSpin}-\eqref{HertzPotential:PosSpin} we get
the master equation \eqref{MasterEq}.

\setcounter{equation}{0}\section{Decoupling limit of near-extreme
Kerr. Mass changing modes}

\label{app:decouple}

In this appendix we show that a decoupling limit of the
near-extremal Kerr black hole yields the NHEK geometry. We follow
\cite{Maldacena:1998uz} where decoupling limits like the one we take
were first discussed for charged non-rotating solutions.

The Kerr black hole solution in Boyer-Lindquist form reads,
\begin{eqnarray}
ds^2&=&\frac{\Sigma \, \Delta }
{\left(\tilde{r}^2+a^2\right)^2-\Delta
a^2\sin^2\theta}\,d\tilde{t}^2 -\frac{\Sigma }{\Delta
}\,d\tilde{r}^2-\Sigma
d \theta^2 \nonumber \\
&& \qquad -\sin^2\theta \frac{\left(\tilde{r}^2+a^2\right)^2-\Delta
a^2\sin^2\theta}{\Sigma } \left(d\tilde{\phi}-\frac{a
\left(a^2+\tilde{r}^2-\Delta
\right)}{\left(\tilde{r}^2+a^2\right)^2-\Delta a^2\sin^2\theta}\,
d\tilde{t}\right)^2, \label{kerr}
\end{eqnarray}
with
\begin{eqnarray}
&& \Delta
=\left(\tilde{r}-\tilde{r}_-\right)\left(\tilde{r}-\tilde{r}_+\right),\qquad
\Sigma
=\tilde{r}^2+a^2\cos^2\theta\,, \nonumber \\
 && \tilde{r}_\pm =\frac{1}{2}\lp \ell_P^2 \Delta E  +\ell_p \sqrt{\ell_P^2 \Delta E^2+ 4J}\rp
\pm \sqrt{\ell_P^3 \Delta E\sqrt{\ell_P^2 \Delta E^2+ 4J}}\,.
\end{eqnarray}
We used the fact that in four dimensions the Planck length $\ell_P$
is related to Newton's constant $G$ by $\ell_P^2=G$, and we defined
the excitation energy above extremality as
 \eq
  \Delta E =\frac{m-a}{\ell_P^2}\,, \qquad {\rm with} \quad
a=\ell_P^2\,\frac{J}{m}\,,
 \eeq
where $J$ and $M=m/\ell_P^2$ are the ADM angular momentum and mass
of the Kerr black hole.

The black hole temperature is
 \eq
 T_H=\frac{r_+^2-a^2}{4 \pi  r_+\left(r_+^2+a^2\right)}\,,
 \eeq
from which follows that near-extremality the relation between the
excitation energy and the temperature is
 \eq
 \Delta E \simeq 8\pi^2J^{\frac{3}{2}}T_H^2\ell_P\,.
 \eeq
Typically, the energy of a quantum of Hawking emission is of order
$T_H$. When this energy is of order or greater than the available
energy above extremality, $T_H\gtrsim \Delta E$, the semiclassical
analysis of the black hole thermodynamics breaks down. This occurs
at an excitation energy of order
 \eq
 E_{\rm gap} \simeq \frac{1}{8\pi^2J^{\frac{3}{2}}\ell_P}\,.
 \eeq

We now want to take a decoupling limit where
 \eq
 \ell_P\rightarrow 0\,, \qquad {\rm with} \quad \lp T_H, J \rp \:\: {\rm
fixed}.\label{DecLim}
 \eeq
In this limit the excitation energy $\Delta E$ vanishes and the gap
energy $E_{\rm gap}$ goes to infinity.

In the decoupling limit, and after introducing the new radial and
azimuthal coordinates $(U,\psi)$:
 \eq
 \tilde{r}=\tilde{r}_+ + 2 \ell_p^2 U \,,\qquad \tilde{\phi} =\psi +\frac{\tilde{t}}{2
m}\,,
 \eeq
the Kerr geometry \eqref{kerr} reduces to
 \eq \frac{ds^2}{\ell_P^2}=
-2J\Omega^2(\theta)
 {\biggl [}-\frac{U(U+4\pi J T_H)}{J^2}\,d\tilde{t}^2+ \frac{dU^2}{U(U+4\pi J T_H)}+
d \theta^2 +\Lambda^2(\theta)\!\lp \!d\psi+\lp 2\pi T_H
+\frac{U}{J}\rp \! d\tilde{t}\rp^2 \!{\biggr ]}, \label{kerrDec}
  \eeq
with $\Omega^2(\theta)$ and $\Lambda^2(\theta)$ defined in
\eqref{NHkerrAux}.

Finally if we introduce the new time, radial and azimuthal
coordinates $(\tau,y,\varphi)$ through the transformations,
\begin{eqnarray}
&& \tilde{t}=\frac{1}{2 \pi  T_H}\left[ \arctan\lp \tau-y \rp+\arctan\lp \tau+y \rp \right] \,, \nonumber \\
&& U=\frac{\pi T_H\,J}{y} \left[ \lp y-1\rp^2-\tau^2\right] \,,\nonumber \\
&& \psi = \varphi + \arctan\lp\frac{\tau - 1}{y}\rp +
\arctan\lp\frac{\tau + 1}{y}\rp\,, \label{kerrDecAux}
\end{eqnarray}
the decoupling geometry \eqref{kerrDec} reduces to
 \eq
\frac{ds^2}{\ell_P^2}= -2J\Omega^2(\theta)
 {\biggl [} \frac{-d\tau^2+dy^2}{y^2}+
d \theta^2 +\Lambda^2(\theta)\lp d\varphi+ \frac{d\tau}{y}  \rp^2
{\biggr ]}. \label{NHkerrPoincare}
 \eeq
We recognize this geometry as the NHEK solution written in
Poincar\'e coordinates. A final transformation between the
Poincar\'e coordinates $(\tau,y,\theta,\varphi)$ and the global
coordinates $(t,r,\theta,\phi)$ \cite{Bardeen:1999px},
\begin{eqnarray}
\tau&=& \frac{\sin t \sqrt{1+r^2} }{\cos t \,\sqrt{1+r^2}+r },\nonumber\\
y&=&\left(\cos t \,\sqrt{1+r^2}+r\right)^{-1},\nonumber\\
\varphi&=&\phi+\ln\left({\cos t+r\sin t \over 1+\sin
t\,\sqrt{1+r^2}}\right), \label{PoincareGlobal}
\end{eqnarray}
 takes
\eqref{NHkerrPoincare} into the NHEK \eqref{NHkerr} written in
global coordinates. Therefore, at the classical level, when we take
the decoupling limit \eqref{DecLim} of the near-extremal Kerr
geometry, one gets a geometry that is independent of the temperature
$T_H$ and that is precisely the NHEK geometry.

An important observation is that going to the next-to-leading order
in the $\ell_P$ expansion, a path similar to
\eqref{DecLim}-\eqref{PoincareGlobal} yields the next-to-leading
order contribution to \eqref{NHkerrPoincare} or \eqref{NHkerr}, that
we call $h_{\mu\nu}$. This $h_{\mu\nu}$ can be considered as a
perturbation to the NHEK since it satisfies the linearized Einstein
equations (with $h\neq 0$) by construction. Asymptotically,
$h_{\mu\nu}$ goes as a power of $r^2$ higher (in all components)
than the GHSS boundary conditions \cite{Guica:2008mu}.

To interpret physically these perturbations we compute the relevant
conserved charges associated to them as defined in
 \eqref{chargeXi}. One finds that the energy of these perturbations is
finite, $Q_{\partial t}[h,g]\neq 0$, while the $U(1)$
charge vanishes, $Q_{\partial\phi}[h,g]=0$. These
perturbations therefore increase the energy of the solution while
leaving its angular momentum unchanged. They correspond thus to
perturbations that take NHEK away from the extremality state. Since
the full series expansion in $\ell_P$ reconstructs near-extreme
Kerr, these perturbations actually take the NHEK geometry into
near-extreme Kerr black hole.


\end{document}